\documentclass[runningheads]{llncs}

\usepackage{amsmath,amssymb}
\usepackage{marvosym}
\usepackage{booktabs}
\usepackage{url}
\usepackage{xcolor}
\usepackage{xspace}
\usepackage{enumitem}
\usepackage{graphicx}
\usepackage{float}
\usepackage{placeins}
\usepackage{subcaption}
\usepackage[ruled,vlined]{algorithm2e}
\SetKwProg{Fn}{Function}{:}{end}
\usepackage{tikz}
\usetikzlibrary{arrows.meta,positioning,calc,patterns}

\makeatletter\AtBeginDocument{\def\doi#1{\url{https://doi.org/#1}}}\makeatother
\newcommand{\findingbox}[1]{\par\smallskip\noindent\fcolorbox{black!35}{black!4}{%
  \parbox{\dimexpr\linewidth-2\fboxsep-2\fboxrule\relax}{#1}}\par\smallskip}
\newcommand{\examplebox}[2]{\par\smallskip\noindent{\setlength{\fboxrule}{0.6pt}\fbox{%
  \parbox{\dimexpr\linewidth-2\fboxsep-2\fboxrule\relax}{\textbf{Running example (#1).} #2}}}\par\smallskip}
\newcommand{\NR}{\textsc{Sluice}\xspace}
\newcommand{\PG}{$\Pi_{\mathrm{glob}}$\xspace}
\newcommand{\PP}{$\Pi_{\mathrm{part}}$\xspace}

\begin{document}

\title{\textsc{Sluice}: Global Invariant, Local Enforcement\\
for Pooled Payment-Channel Liquidity}
\titlerunning{\textsc{Sluice}: Global Invariant, Local Enforcement}
\author{Yueqi Wu\inst{1} \and Huiping Sun\inst{1}\textsuperscript{(\Letter)} \and Peilu Guo\inst{1} \and
Yiming Zhu\inst{2} \and Zhong Chen\inst{2,3}}
\authorrunning{Y. Wu et al.}
\institute{School of Software \& Microelectronics, Peking University, Beijing, China\\
\email{wuyueqi77@stu.pku.edu.cn, sunhp@ss.pku.edu.cn, gpeilu@stu.pku.edu.cn}
\and
School of Computer Science, Peking University, Beijing, China\\
\email{ymzhu25@stu.pku.edu.cn, zhongchen@pku.edu.cn}
\and
School of AI and Liberal Art, Beijing Normal-Hong Kong Baptist University,
Zhuhai, China}

\maketitle

\begin{abstract}
A routing node on the Lightning Network holds its liquidity in separate
channels, so a payment can fail at a channel whose outbound balance is
exhausted while the node's other channels still hold balance. Pooling the
channels into one reserve fixes this only if the node's draws across all
channels stay within the reserve: a global invariant that each counterparty
must enforce from its own channel, with no shared counter that off-chain
draws can update. A node must therefore split the reserve into per-channel
quotas in advance or coordinate every draw with every counterparty. On three
Lightning snapshots the advance split forfeits $16$ to $67\%$ of the pooling
gain over unpooled channels, and the loss grows with channel count. \NR
recovers $19$ to $65\%$ of that loss with a nested reservation: each channel
keeps an exclusive base that its counterparty checks alone, and the rest is
a shared overflow drawn on with certificates from a capacity-weighted quorum
of the node's counterparties. Two conflicting certificates share an honest
signer, so over-drawing is prevented without a slashable stake, under a
stated bound on the capacity the node controls in the signing set. \NR
loses at most $1.3$ points of payment success to coordination where
deployed coin movers lose up to $10.3$, and improves them in eleven of
twelve cells when stacked on them. Re-creating every base output each epoch
costs $2.8$ to $5.1$ times Lightning's on-chain bytes; re-creating only
those that overflowed costs $0.5$ to $1.2$ times and keeps part of the
gain.
\end{abstract}

\section{Introduction}
\label{sec:intro}

A routing node on the Lightning Network holds its capital in separate
channels, and each channel serves one
counterparty~\cite{poon2016bitcoin,gudgeon2020sok}. Liquidity in a channel
is directional: the node's outbound side empties as its inbound side fills,
so sustained one-way flow exhausts the channels one at a time while the others
still hold balance~\cite{van2021merchant,sivaraman2020spider}. The node is
not short of liquidity; its liquidity sits in its other channels, and moving
it costs an on-chain transaction. The natural fix is to let the node post one reserve,
skimmed from the channels it already funded, that backs a draw on any of
them. No capital is added, but the capital the node has becomes fungible
across its channels, and on real Lightning snapshots that fungibility is
worth a large share of payment success (\S\ref{sec:eval}).

\paragraph{Global invariant, local enforcement.}
A draw is safe for the counterparty who accepts it only if the node's draws
across all channels stay within the reserve; otherwise the same coins back
two promises. A shared counter that every draw decrements would settle this,
but off-chain draws have none: on an account chain it costs an on-chain write
per draw, and on Bitcoin an update signed by all $n$ counterparties. Each
counterparty sees only its own channel, and the node cannot be asked, since
it is the party that gains from over-drawing. The invariant is global and
its enforcement local: each counterparty decides from its own view and,
beyond its base, from a quorum's certificate. We call this problem
\emph{global invariant, local enforcement}.

\begin{figure}[t]
\centering
\begin{tikzpicture}[font=\footnotesize,x=0.95cm,y=0.75cm,
  ax/.style={-{Stealth[length=1.6mm]},thin,black!70},
  pt/.style={circle,fill,inner sep=1.6pt},
  cpt/.style={font=\scriptsize\itshape,black!55},
  meas/.style={font=\scriptsize,black!60,align=left}]
\draw[ax] (0,0) -- (9.0,0);
\node[font=\scriptsize,anchor=north east] at (9.0,-0.08) {coordination per draw};
\draw[ax] (0,0) -- (0,4.6);
\node[font=\scriptsize,anchor=north west,rotate=90] at (0.1,0.55) {liquidity made fungible};
\draw[dashed,black!45] (0.2,0.25) -- (8.8,0.25);
\node[meas,anchor=south east] at (8.8,0.27) {LN, no pool: $45.4\%$};
\draw[very thick,black!75] plot[smooth,tension=0.75]
  coordinates {(0.7,1.65) (2.2,2.8) (3.8,3.1) (5.6,3.22) (7.6,3.3)};
\node[pt] at (0.7,1.65) {};
\node[anchor=north west,align=left] at (0.85,1.5) {\textbf{Partition} $\Pi_{\mathrm{part}}$};
\node[anchor=north west,cpt] at (0.88,1.08) {reserve everything in advance};
\node[meas,anchor=north west] at (0.88,0.72) {$57.5\%$, no per-draw coordination};
\node[pt] at (2.2,2.8) {};
\node[anchor=north west,align=left] at (2.5,2.72) {\textbf{\NR} $\Pi(\alpha)$};
\node[anchor=north west,cpt] at (2.52,2.3) {exclusive base, quorum-certified overflow};
\node[meas,anchor=north west] at (2.52,1.94) {$63.2\%$, $39\%$ of draws reach a quorum};
\node[pt] at (7.6,3.3) {};
\node[anchor=south east,align=right] at (7.7,4.15) {\textbf{Global check} $\Pi_{\mathrm{glob}}$};
\node[anchor=south east,cpt,align=right] at (7.7,3.8) {an on-chain write or $n$-of-$n$ per draw};
\node[meas,anchor=south east,align=right] at (7.7,3.42) {$67.6\%$, every draw reaches all $n$};
\end{tikzpicture}
\caption{The idea of the paper. Enforcing a global bound from local state
costs either advance reservation or per-draw coordination. The curve is
schematic; the four values are payment success measured on the 2021
snapshot under skew at $L=20$ (Table~\ref{tab:main}), the \NR point at
$\alpha=0.5$. The global check is an upper bound, not a design. One knob
$\alpha$ moves \NR along the curve.}
\label{fig:thesis}
\end{figure}
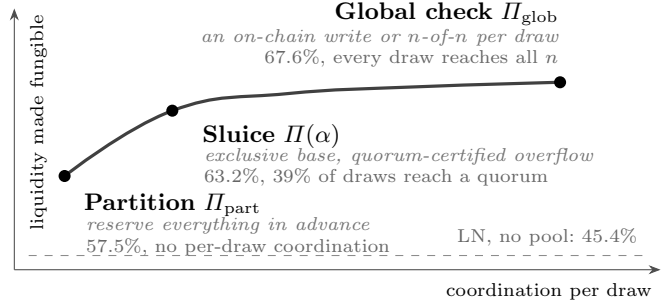

\paragraph{Two enforcement endpoints.}
Enforcing one bound from many local views has two endpoints
(Fig.~\ref{fig:thesis}). A node can reserve in advance, splitting the
reserve into per-channel quotas that each counterparty checks alone; this
needs no per-draw coordination but commits capacity before demand is known.
Or it can check every draw against one shared counter, which exists only as
an on-chain write per draw or an $n$-of-$n$ update, an upper bound rather
than a design. On three Lightning snapshots the advance split forfeits $16$
to $67\%$ of what pooling adds, and the loss grows with channel count
(\S\ref{sec:pooling}).

\paragraph{Sluice.}
Revenue management met the same problem with airline seats and
\emph{nests} its booking limits, so that capacity reserved for one fare
class stays reachable by another~\cite{talluri2004theory}. \NR, named for
the gate that lets water pass between sealed compartments, adapts this to
an untrusted seller and distributed checks. Each channel keeps an exclusive
base that its counterparty checks alone, exactly as today, and the rest of
the reserve is a shared overflow that any channel may draw on with a
certificate from a capacity-weighted quorum of the node's counterparties.
Two conflicting certificates must share an honest signer, who signs at most
one, so over-drawing is prevented rather than punished; the common case
sends no message, and a node whose quorum is absent falls back to its
bases. One knob $\alpha$ sets the split: $\alpha=1$ is the partition,
$\alpha=0$ shares everything.

The assumption specific to \NR, stated up front and not enforced, is that
the node controls at most a bounded share of the signing set's announced
capacity; a sybil channel counts only for what the node actually locked in
it. Prior reservation schemes assume a trusted
allocator~\cite{oneil1986escrow,barbara1994demarcation,balegas2015bounded}
and prior quorum schemes a shared replicated
state~\cite{malkhi1998byzantine}; \NR handles a Byzantine principal,
checkers that each see one channel, and no shared counter.

\paragraph{Contributions.}
\begin{enumerate}[leftmargin=*,nosep]
\item \textbf{We measure what reserving in advance costs.} Against an
idealised global check at equal capital, with coordination charged to every
deployable mechanism under one rule, we measure what the split forfeits on
three Lightning snapshots and trace the loss to channel count
(\S\ref{sec:pooling}, \S\ref{sec:eval-q1}, \S\ref{sec:eval-q2}).
\item \textbf{We give \NR, a nested reservation protocol for Bitcoin.} Its
bound holds by quorum intersection rather than punishment; it uses deployed
Bitcoin script, falls back safely when the quorum is absent, and is
specified down to scripts, per-party state and settlement, with a regtest
prototype of the anchor and its re-anchoring (\S\ref{sec:nest},
\S\ref{sec:security}).
\item \textbf{We evaluate it against deployed alternatives and an
adversary.} \NR recovers $19$ to $65\%$ of the forfeited gain, loses at most
$1.3$ points to coordination where Shaduf and Horcrux lose up to $10.3$,
stacks positively on both in eleven of twelve cells, by up to $8.3$ points,
and holds its bound in every attack on the specified protocol, at $2.8$ to $5.1$ times Lightning's on-chain bytes
with full anchors or $0.5$ to $1.2$ times with incremental ones
(\S\ref{sec:eval}).
\end{enumerate}

\section{Background}
\label{sec:background}

A payment channel locks capacity between two parties, who then pay each
other by signing new balance splits off-chain and settle on-chain only to
open or close~\cite{poon2016bitcoin,dziembowski2018general,%
aumayr2021generalized}. A network routes a payment over a path of channels
with a hash-locked contract at each hop, so every hop moves balance one
way~\cite{malavolta2017concurrency,roos2018speedymurmurs,gudgeon2020sok}, and
sustained one-way demand leaves a routing node with balance in channels that
do not need it and none in those that do~\cite{van2021merchant,sivaraman2020spider}. The deployed remedies
relocate capital between channels: cycle
rebalancing~\cite{khalil2017revive}, coin-shifting between a node's own
channels~\cite{ge2022shaduf,shadufpp,avarikioti2022hide,aumayr2022thora},
pooling across neighbours~\cite{tian2024horcrux,starfish}, and channel
factories~\cite{burchert2018factories}; Table~\ref{tab:related} sets out what
each must coordinate. Buying inbound capacity~\cite{lspspec} trusts the
seller, and splicing fresh funds in is the on-chain remedy our evaluation
counts as forced refills. Admission control~\cite{bastankhah2023r2} and
topology optimisation~\cite{chatterjee2025topology} choose which draws to
accept. All of these move or ration liquidity; we ask how a node makes the
liquidity it already owns fungible across its channels, which is orthogonal
to moving it and, as \S\ref{sec:eval-q1} shows, composes with it.

\begin{table}[t]
\centering\footnotesize
\setlength{\tabcolsep}{4pt}
\caption{Where this work sits. Coin movers relocate capital;
\NR enforces a shared reserve.}
\label{tab:related}
\begin{tabular}{@{}llll@{}}
\toprule
 & moves coins & who must agree & shared reserve \\
\midrule
Revive~\cite{khalil2017revive}      & yes & a routable cycle       & no  \\
Shaduf~\cite{ge2022shaduf}          & yes & the two channel ends   & no  \\
Horcrux~\cite{tian2024horcrux}      & yes & neighbours in a group  & partial \\
Factories~\cite{burchert2018factories} & yes & all members, fixed set & no \\
\midrule
$\Pi_{\mathrm{part}}$ (baseline)    & no  & nobody                 & yes, pre-split \\
\NR (this work)                     & no  & a quorum, on overflow only & yes, nested \\
\bottomrule
\end{tabular}
\end{table}

Keeping a global invariant without coordinating on every operation is an
old problem: escrow transactions~\cite{oneil1986escrow}, the demarcation
protocol~\cite{barbara1994demarcation} and bounded-counter
CRDTs~\cite{balegas2015bounded} give each site a reservation whose sum cannot
violate the bound. Our baseline is that construction with distrustful sites
and a Byzantine principal, and uniqueness from Bitcoin's double-spend
rule~\cite{nakamoto2008bitcoin,todd2016seals}. Nesting booking limits is revenue
management's answer to a wasteful
partition~\cite{talluri2004theory,eppen1979effects,paterson2011inventory};
its safety here is a quorum-intersection
argument~\cite{malkhi1998byzantine}. What the literature does not supply is
a measurement of what the reservation costs.

\section{Global Invariant, Local Enforcement}
\label{sec:pooling}

\paragraph{Pooling a node's liquidity.}
A node $N$ has channels $1,\dots,n$ with counterparties $w_1,\dots,w_n$.
Today a payment that needs more than channel $i$ holds fails at this hop even
when the other channels still hold balance. Pooling changes the accounting: the node
skims a fraction $\varphi$ of its balance from every channel into one reserve
$B$, and a shortfall on channel $i$ can be met by drawing $\delta$ from $B$.
Write $D_i$ for the total drawn on channel $i$. The reserve is credible only
if
\begin{equation}
\textstyle\sum_i D_i \;\le\; B ,
\label{eq:bound}
\end{equation}
since otherwise the same coins back two promises and some counterparty is
unbacked. \eqref{eq:bound} is a global invariant, but each $w_i$ observes only channel $i$ and must decide, from
that alone, whether accepting a draw keeps \eqref{eq:bound} true.

\paragraph{Two endpoints.}
\label{sec:space}
The \emph{global check} \PG tests every draw against the node's current
total, $\delta \le B - \sum_j D_j$. It strands nothing and needs a shared
variable written on every draw; on Bitcoin that is an $n$-of-$n$ off-chain
state every counterparty updates every time. We use \PG as the upper bound on
pooling the node's \emph{own} reserve, not as a proposal and not as a bound
on liquidity management at large, since a coin mover also reaches the
counterparty's balance and can exceed it (Appendix~\ref{app:ideal}). The
\emph{partition} \PP splits the reserve each epoch into per-channel quotas
$\ell_i$ with $\sum_i \ell_i \le B$, commits them under one anchored root,
and lets $w_i$ accept draws while $D_i \le \ell_i$, talking to nobody. It is
safe by construction and needs no per-draw coordination, so it is our baseline; its price
is that capacity is committed before anyone knows where it will be needed.

\paragraph{Why the price grows with channel count.}
\begin{proposition}[Reservation fragmentation]
\label{prop:frag}
Under an even split of a reserve $B = \varphi n \bar b$, where $\bar b$ is
the mean channel balance, a channel can draw at most $\ell = \varphi \bar b$,
while under \PG it can draw up to $B$; the ratio is $n$.
\end{proposition}
The share $\ell$ does not shrink as channels are added, since the reserve is
skimmed from all of them, but the ceiling one channel could reach grows with
every channel added, so the capacity a partition strands is largest on the
nodes with the most channels, which are the nodes that route. This is a
bound on what reservation can strand, not a prediction of payment success,
and for any allocator some channel gets at most $B/n$, so the ratio is at
least $n$; \S\ref{sec:eval-q2} measures it node by node. Re-anchoring more often, a larger reserve and allocation by
predicted demand each leave most of the gap (Appendix~\ref{app:space}), and
\emph{optimistic accounting}, which punishes over-borrowing afterwards, is
unsafe. The remedy must be structural.

\examplebox{1/3}{A node has $n=4$ channels of equal capacity and a
reserve $B=16$. The partition gives each channel $\ell=4$, so a payment
that needs channel~1 to draw $8$ fails at this hop although nothing has been
drawn anywhere; \PG would let it reach all $16$. Under optimistic
accounting the node tells each counterparty it has drawn nothing elsewhere
and draws $16$ on every channel, $64$ against a reserve of $16$.}

\section{\NR}
\label{sec:nest}

Each epoch the node anchors every channel's exclusive base and one shared
overflow on chain. A draw within the base is checked by the counterparty
alone; a draw beyond it needs one round of signatures from a
capacity-weighted quorum, chained so that the overflow is consumed once.
Appendix~\ref{app:ds} has the notation, the certificate format, the
pseudocode and every failure case.

\subsection{The anchor and the allocation}
\label{sec:nest-epoch}
\label{sec:nest-alloc}
Time is divided into epochs, a re-allocation cadence rather than a safety
parameter. At the start of epoch $t$ the node publishes one \emph{anchor
transaction} with four kinds of output:
\begin{itemize}[leftmargin=1.2em,nosep]
\item $n$ \emph{base outputs}, output $i$ holding $\mathrm{base}_i$. It is
spent by the node and $w_i$ together (a re-anchor or a cooperative close),
by $w_i$ alone after a timelock $T_1$ (the node has stopped cooperating), or
by the node alone after a longer $T_2$ ($w_i$ has gone for good);
\item one \emph{overflow output}, spent by a quorum of the signing set
$S_t$, or by the node alone after a timelock $\Delta'$;
\item a \emph{thread output} that the next anchor must spend; and
\item an \texttt{OP\_RETURN} carrying a commitment $R_t$, a Merkle sum tree
over the leaves $\langle i, \mathrm{base}_i\rangle$ whose root also fixes
the total.
\end{itemize}
No output has a key path that one party can use alone. A re-anchor spends
the previous thread, the previous overflow with the quorum's signatures, and
each base it re-creates with its counterparty's signature; a counterparty
that does not answer leaves its base standing until the next anchor. Each
$w_i$ receives its leaf, its path and the overflow's size. Two properties
do the work in \S\ref{sec:security}: the anchor spends the previous thread,
so a second root in one epoch would be a double-spend; and the bases are
separate outputs, so on a simultaneous default every counterparty claims its
own without racing the others.

Let $\mathrm{free} = \max(0,\,B - \sum_i D_i)$ be the capacity the reserve
still allows. Then
\begin{align}
\mathrm{base}_i &= D_i + \alpha\,\mathrm{free}/n, \label{eq:base}\\
\mathrm{overflow} &= (1-\alpha)\,\mathrm{free}. \label{eq:over}
\end{align}
Debt is carried: each base starts at the channel's outstanding $D_i$, so an
epoch boundary never invalidates a draw already accepted, and new capacity
comes only from $\mathrm{free}$. The split of $\mathrm{free}$ is even on
purpose, since the demand-predicting rules we tried did worse in five of
six cells. A base
whose channel did not overflow still covers its debt and can stand, so the
node may instead re-create only the overflow, the thread and the bases of
channels that overflowed; \S\ref{sec:eval-q2} prices this incremental form
and Appendix~\ref{app:fail} gives its allocation.

\subsection{Draws}
\label{sec:nest-draw}
Fig.~\ref{fig:mech} shows the two paths and Algorithm~\ref{alg:draw} in
Appendix~\ref{app:ds} states them line by line.

\begin{figure}[t]
\centering
\begin{tikzpicture}[
  font=\scriptsize,
  box/.style={draw,rounded corners=1pt,minimum height=3.6mm,inner sep=2pt,
              align=center},
  oput/.style={box,fill=black!4,minimum width=15mm},
  lbl/.style={font=\scriptsize\itshape},
  ar/.style={-{Stealth[length=1.4mm]},thin},
]
\node[lbl,anchor=west] at (-1.3,2.5) {(a) anchor transaction, epoch $t$};
\node[box,minimum width=20mm] (thr) at (-0.2,1.75) {thread $t{-}1$};
\node[box,minimum width=20mm] (res) at (-0.2,1.05) {overflow, bases $t{-}1$};
\node[draw,minimum width=4mm,minimum height=22mm,fill=black!8]
  (tx) at (1.55,0.85) {};
\node[rotate=90,font=\scriptsize] at (1.55,0.85) {tx$_t$};
\draw[ar] (thr) -- (thr -| tx.west);
\draw[ar] (res) -- (res -| tx.west);

\node[oput] (b1) at (3.65,1.95) {$\mathrm{base}_1$};
\node[oput] (b2) at (3.65,1.42) {$\cdots$};
\node[oput] (bn) at (3.65,0.89) {$\mathrm{base}_n$};
\node[oput] (ov) at (3.65,0.30) {overflow};
\node[oput,minimum width=22mm] (op) at (3.95,-0.29)
  {thread, \texttt{OP\_RETURN} $R_t$};
\foreach \n in {b1,b2,bn,ov,op} \draw[ar] (tx.east) -- (\n.west);
\node[anchor=west,align=left,font=\scriptsize] (n1) at (5.55,1.42)
  {distinct UTXOs,\\ all $n$ claims confirm};
\draw[ar,black!55] (b1.east|-n1) -- (n1.west);
\node[anchor=west,align=left,font=\scriptsize] (n2) at (5.55,0.30)
  {quorum of $S_t$,\\ or $N$ after $\Delta'$};
\draw[ar,black!55] (ov.east) -- (n2.west);

\begin{scope}[yshift=-1.35cm]
\node[lbl,anchor=west] at (-1.3,0.45) {(b) a draw of $\delta$ on channel $i$};
\node[box] (q) at (0.0,-0.35) {$D_i{+}\delta \le \mathrm{base}_i$?};
\node[box,fill=black!4,anchor=west] (loc) at (2.4,0.0)
  {accept locally against $R_t$, no messages};
\node[box,fill=black!4,anchor=west] (quo) at (2.4,-0.72)
  {collect a quorum on $(R_t,\delta',C',h_{\mathrm{prev}},b)$};
\draw[ar] (q.east) -- ++(0.35,0) |- node[pos=0.72,above,font=\scriptsize]
  {yes} (loc.west);
\draw[ar] (q.east) -- ++(0.35,0) |- node[pos=0.72,below,font=\scriptsize]
  {no} (quo.west);
\node[anchor=west,font=\scriptsize] at (2.4,-1.25)
  {quorum unreachable, so fall back to $\mathrm{base}_i$
   (Prop.~\ref{prop:failsafe})};
\end{scope}
\end{tikzpicture}
\caption{\NR in one epoch. The base path sends no message; the overflow path
needs a quorum, and is the only place counterparties must cooperate.}
\label{fig:mech}
\end{figure}
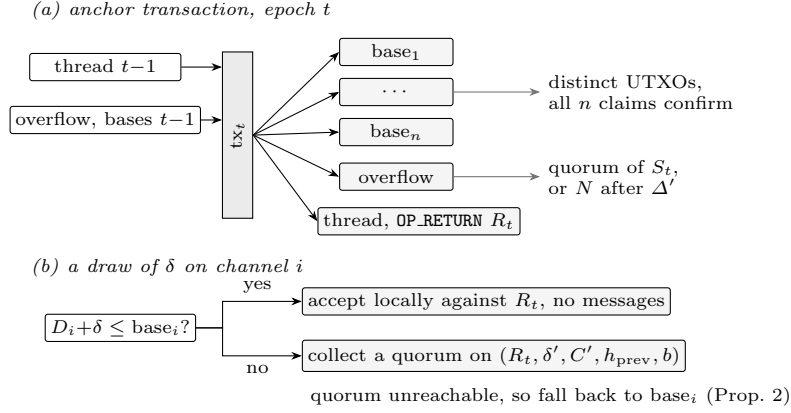

\begin{description}[leftmargin=1.2em,itemsep=2pt]
\item[Base path.] $w_i$ accepts $\delta$ when $D_i+\delta \le \mathrm{base}_i$,
checking its own leaf against $R_t$ and nothing else. No message leaves the
channel. This is \PP, and it carries the common case: at $\alpha=0.9$ only
$9$ to $29\%$ of draws ever leave it.
\item[Overflow path.] When the draw exceeds the base by $\delta'$, the node
presents a certificate on $(R_t, \delta', C', h_{\mathrm{prev}}, b)$ signed
by a quorum of $S_t$: $C'$ is the cumulative overflow consumed \emph{after}
this draw, $h_{\mathrm{prev}}$ the hash of the certificate that set the
previous total, and $b = h(i \,\|\, r)$ binds it to channel $i$ under a salt
$r$ only $w_i$ receives. $w_i$ accepts only if the signatures reach the quorum
and $b$ opens to its $i$, so one certificate serves one channel; the signers
see $b$, not $i$.
\end{description}

\paragraph{What a signer checks.}
A signer $w_j$ keeps three durable values: $c_j$, the last certified total
it knows; $h_j$, the hash of the certificate that set it; and $p_j$, the
request it has signed on top of $h_j$ and not yet seen certified. It signs
a request only if
\begin{enumerate}[leftmargin=2em,nosep,label=(\arabic*)]
\item the request names this epoch's root $R_t$;
\item $C' \le \mathrm{overflow}$, so the overflow is never exceeded;
\item $C' = c_j + \delta'$, so the certificate accounts for exactly what it
authorises;
\item $h_{\mathrm{prev}} = h_j$, so it extends the one chain $w_j$ has
endorsed; and
\item $p_j$ is empty or equals this request, with $p_j$ written to disk
before the signature leaves.
\end{enumerate}
It advances $(c_j, h_j)$ and clears $p_j$ when shown a certificate with a
valid quorum that extends $h_j$, whether or not it signed it. Check (5) is
what turns quorum intersection into a global bound: two conflicting
successors of one state would need two quorums sharing an honest signer, who
signs at most one. Certificates form a chain, so the node collects one at a
time; a draw that arrives while another is in flight falls back to its base,
a cost the evaluation charges.

\examplebox{2/3}{At $\alpha=0.5$ the anchor commits $\mathrm{base}_i=2$ for every channel and
an overflow of $8$. Channel~1 draws $2$ from its base with no message and
$6$ beyond it with a certificate for $C'=6\le 8$; with the four members'
capacities normalised to $1$ each and a node assumed to control at most $F=1$ of it, the
certificate needs signers holding more than $(4+1)/2$, any three of the
four (\S\ref{sec:signers}). A later request of $5$ on channel~2 would need
$C'=9>8$, so no honest signer signs it; the draw falls back to the base of
$2$ and the payment fails at this hop, as it does today. At the next anchor
$D=(8,0,0,0)$ and $\mathrm{free}=8$, so the node commits bases
$(9,1,1,1)$ and an overflow of $4$: channel~1's debt is carried and only
the free $8$ is re-split.}

\subsection{The signing set}
\label{sec:signers}
The overflow is only as safe as the quorum, and the node chooses who signs,
so membership is verifiable and sybil membership costs capital.
$S_t$ is the set of the node's counterparties that opted in by a feature bit
and whose channel is announced in gossip with capacity at least $c_{\min}$.
The anchor's tapleaf lists their keys, so every counterparty can read $S_t$,
check each member against gossip, and rely on it not changing within the
epoch.

\paragraph{Capacity-weighted quorum.}
Let $\mathrm{cap}_j$ be member $j$'s announced capacity, which is the value
of the funding output the announcement names and so checkable on chain,
and $W$ their sum. A certificate is valid when its signers' capacity
exceeds $(W + F)/2$, where $F$ is the capacity the accepting counterparty
assumes the node controls within $S_t$ (Lemma~\ref{lem:wqi}). A sybil channel
counts only for what the node locked in it: $20\%$ of the capacity of the
median hub (a node with more than $32$ channels)
capacity is $0.3$ to $0.8$\,BTC on our snapshots. $F$ must over-estimate. A
counterparty that under-estimates it loses the guarantee, one that
over-estimates it loses only availability, and one that believes the node
controls a third of $S_t$ or more should decline overflow certificates.
Appendix~\ref{app:signers} gives the workable range of $F$, counterparties
with different $F$, the membership floor $c_{\min}$ and its price, and how
the on-chain count threshold relates to the weight.

\paragraph{Why counterparties sign.}
A member that withholds signatures can be dropped from $S_t$ at the next
anchor, and a node may make membership the price of drawing on the
overflow (our simulator lets every channel draw). The node also pays each
signer a small fee over the channel it already has with it; one satoshi per
signature is under a third of a typical forwarding fee on a
$10{,}000$-satoshi payment. The fee is part of the design, since most nodes
never draw (\S\ref{sec:eval-q2}).

\subsection{Binding \NR to Bitcoin and Lightning}
\label{sec:deploy}
\label{sec:instantiation}
\NR is an enforcement layer one level above Lightning's channel protocol;
we claim it as Bitcoin-realisable rather than deployed.

\paragraph{Scripts and settlement.}
The overflow protocol is entirely off-chain: signatures over the certificate
tuple, no covenant, no consensus change. Every anchor output is a Taproot
output whose internal key is the unspendable point of BIP~341, so it is
spent only through the script paths of \S\ref{sec:nest-epoch}: the base's
cooperative path is a 2-of-2 of the node and $w_i$ (a MuSig2 key in
deployment), and the overflow's quorum path a $k$-of-$m$ tapleaf. The
tapleaf enforces the signature threshold and nothing else; the semantic
checks live in the signers, who check a default settlement against the
certificate chain once the creditors open their salts
(Appendix~\ref{app:fail}). A regtest prototype builds and checks a full
re-anchor; a certificate costs $0.5$\,ms (Appendix~\ref{app:proto}).

\paragraph{Binding a draw to the channel.}
A draw on channel $i$ is an HTLC forward in which the node's balance in the
commitment goes below zero by up to $\mathrm{base}_i$ on the base path, and
by the certified $\delta'$ beyond it on the overflow path. Admitting that
negative balance is the one change \NR needs in the commitment format: the
commitment carries a credit field, bounded by $\mathrm{base}_i$ plus the
certified overflow. The deficit is paid from the base output through its
cooperative path; if the node stops cooperating, $w_i$ takes the whole base
output after $T_1$, so an unresponsive node forfeits its headroom. A
channel whose peer lacks the feature bit runs at credit zero, which is
Lightning today. $w_i$ checks the leaf, or the certificate, before it sends
\texttt{commitment\_signed}.

\paragraph{Latency, lifecycle and messages.}
A round trip is about $0.2$ to $0.5$\,s, the order of one forwarding hop;
$L=20$ payment arrivals is one round trip for a hub at $100$ payments per
second. A channel closed or opened
mid-epoch takes effect at the next anchor (Appendix~\ref{app:fail}). \NR adds
three messages, for the leaf and path, the request and the signature, and a
feature bit for opting into $S_t$.

\section{Security}
\label{sec:security}

\paragraph{Assumptions.}
The node is arbitrarily Byzantine: it may propose any draw, show different
messages to different counterparties, withhold and reorder, but cannot forge
signatures or change an anchored root. Within $S_t$ the
node controls at most $F$ of announced capacity, including channels it opened
to itself, and with $c_{\min} > 2F/m$ fewer than $k$ members. This is
assumed, not enforced: each counterparty chooses its own $F$, and one that
under-estimates it loses P1. Honest counterparties follow the protocol, keep
$(c_j, h_j, p_j)$ and their salts durably, and may be offline; anchors
confirm within $\Delta$, well inside the timelocks $T_1$, $T_2$ and
$\Delta'$.
Liability safety does not rely on signer rationality; availability and
participation do.

\paragraph{What is guaranteed.}
Four properties, kept apart because they rest on different assumptions:
\begin{enumerate}[leftmargin=*,nosep]
\item[P1] \emph{Liability safety}: $\sum_i D_i \le B$ at all times.
\item[P2] \emph{Claim uniqueness}: each certificate serves exactly one channel.
\item[P3] \emph{Recoverability}: on default, every counterparty recovers its
base alone, and overflow creditors recover their certified claims when $k$
members willing to sign are reachable: the node holds fewer than $k$, so any
$k$ include an honest signer, who signs only the chain-determined split, and
if the node's members refuse, $k$ honest members must be reachable.
\item[P4] \emph{Fail-safe degradation}: when no quorum is reachable, no new
overflow draw is accepted; the epoch's bases stand, and the node restores
\PP by anchoring with $\alpha=1$ once the bases' counterparties co-sign or
their timelocks expire.
\end{enumerate}
P1 is safety and rests on the capital bound and durable signer state; P2
rests on the salted binding.
P3 is the one place where a liveness failure becomes an exposure: if the
timelock returns the overflow output to the node, overflow creditors lose
their overflow claims and nothing else. P4 needs no assumption for safety.

\begin{lemma}[Endpoint, $\alpha=1$]
\label{thm:base}
At $\alpha = 1$ each $w_i$ accepts at most $\mathrm{base}_i$, every epoch
maintains $\sum_i \mathrm{base}_i = \max(B, \sum_i D_i)$, and new capacity
comes only from $\mathrm{free}$; hence \eqref{eq:bound} holds at every later
time from any state satisfying it.
\end{lemma}
\begin{lemma}[Weighted quorum intersection]
\label{lem:wqi}
If two signer sets carry capacity above $(W+F_i)/2$ and $(W+F_j)/2$, their
intersection carries capacity above $(F_i+F_j)/2$; hence if the node
controls at most $\min(F_i,F_j)$, the intersection contains an honest signer.
\end{lemma}
\begin{lemma}[Certificate uniqueness]
\label{lem:unique}
Under the capital bound, every certified state has at most one certified
successor, so the certificates of an epoch form a single chain.
\end{lemma}
\begin{theorem}[Overflow soundness]
\label{thm:overflow}
For every $\alpha \in [0,1]$ the total consumed from the overflow never
exceeds $\mathrm{overflow}$, and therefore \eqref{eq:bound} holds.
\end{theorem}
\begin{proposition}[Fail-safe degradation]
\label{prop:failsafe}
If the node withholds the overflow state, or no quorum of $S_t$ is
reachable, no certificate forms and every counterparty falls back to its
base for the epoch; within the epoch liveness loss never becomes safety loss
(the exposure at settlement is P3's), though the base
alone is less than the partition would have given until the node re-anchors
with $\alpha=1$.
\end{proposition}
Proofs are in Appendix~\ref{app:proofs}. They rest on two facts: two valid
quorums share an honest signer, and an honest signer signs at most one
successor of any state because $p_j$ is persisted before its signature
leaves. Table~\ref{tab:assume} lists what each guarantee rests on.

\begin{table}[t]
\centering\footnotesize
\setlength{\tabcolsep}{5pt}
\caption{Assumptions and what they buy.}
\label{tab:assume}
\begin{tabular}{@{}ll@{}}
\toprule
assumption & guarantee \\
\midrule
Bitcoin consensus & one root per epoch \\
node holds $\le F$ of $S_t$'s capacity & no conflicting certificates (P1) \\
$w_i$ keeps its salt $r$ & one channel per certificate (P2) \\
node holds $< k$ members & node cannot assemble settlement (P3) \\
signers persist $(c_j,h_j,p_j)$ & one successor per state (P1) \\
quorum reachable at settlement & overflow claims recovered (P3) \\
none & base claims recovered (P3) \\
co-signature or timelock & fallback to \PP (P4) \\
$F$ under-estimated & \emph{P1 lost} \\
\bottomrule
\end{tabular}
\end{table}

\examplebox{3/3}{Suppose the node shows $\{w_1,w_2,w_3\}$ a request for $C'=6$ on the
epoch's first state and $\{w_2,w_3,w_4\}$ a second request for $C'=6$ on the
same state, hoping to draw $12$ from an overflow of $8$. The two sets share
two members, at most one the node's, so an honest one has recorded the first
request in $p_j$ and refuses the second. If the node controls $w_3$ and
$w_2$ forgot the first after a crash, the restored $w_2$, $w_3$ and $w_4$
would sign the second; this is why a signer that lost its durable state
signs nothing until the next anchor (Appendix~\ref{app:fail}).}

\section{Evaluation}
\label{sec:eval}

\findingbox{\textbf{Evaluation questions.}
\textbf{Q1 Effect}: what does pooling achieve against what a node can
deploy today, and does \NR compose with it?
\textbf{Q2 Cost}: what does \NR cost in coordination, signers'
cooperation and on-chain bytes?
\textbf{Q3 Safety}: does the bound hold against an adversary running
against the implementation?
\textbf{Q4 Parameters}: how do the results move with $\alpha$ and
$\varphi$?}

\subsection{Setup}
\label{sec:setup}
\emph{Snapshots.} Three Lightning graphs:
\textbf{2021} ($10{,}529$ nodes, $38{,}910$ channels), the one the released
Shaduf and Horcrux simulators ship, so the baselines run on their own data;
\textbf{2023} ($15{,}071$ nodes, $64{,}194$ channels), the last in a
published archive~\cite{valko2025geo}; and \textbf{2026} ($5{,}507$ nodes,
$27{,}063$ channels), collected on 4 August 2026 from Rapid Gossip
Sync~\cite{rgs2024,lightning-rfc}, whose capacities are announced HTLC
maxima, so it serves for its concentration rather than as a liquidity
measurement (Appendix~\ref{app:sens} perturbs it).
\emph{Workloads.} Payment amounts come from a released March 2021 on-chain
sample; \emph{skew} selects senders with skew
parameter $4$, \emph{drift} sends one-directional hub traffic at four times
the observed amounts, and a plain replay appears in
Appendix~\ref{app:sens}. Topology is fixed within a run.
\emph{Parameters.} Every channel's capacity is scaled by a capacity factor,
$4$ unless stated. Arms that skim use $\varphi=0.30$ and re-anchor every
$10{,}000$ payments (the on-chain sweep of \S\ref{sec:eval-q2} also uses
$3{,}500$, $14{,}000$ and $56{,}000$); runs are $50{,}000$ measured payments,
ten seeds in the main table and five elsewhere, with $95\%$ $t$ intervals.
\emph{Baselines.} Shaduf and Horcrux are migrated from the released
simulators and reproduce their reported ratios within $4\%$
(Appendix~\ref{app:fidelity}).
\emph{Artifact.} The simulator, every experiment script, the raw results
and the prototypes will be made public upon acceptance.

\emph{Coordination.} Prior comparisons run every mechanism at zero
coordination latency; we charge every arm under one rule, in which an
attempt succeeds only if the node's coordination slot is free and the
parties it must reach answer, each with probability $p$, within a window of
$L$ payment arrivals that we sweep. Arms differ only in whom they must
reach: Horcrux the upstream neighbour, Shaduf another channel's peer,
\NR's overflow any quorum of $S_t$, and \NR's base path and LN nobody
beyond the next hop; \PG is not charged (Appendix~\ref{app:attacks}).

\subsection{Q1: what pooling achieves}
\label{sec:eval-q1}

\begin{table}[t]
\centering\footnotesize
\setlength{\tabcolsep}{3.5pt}
\caption{Payment success at capacity factor $4$ with coordination charged
($L{=}20$, $p{=}0.9$, $\varphi{=}0.30$ for every pooled arm), ten seeds,
$95\%$ half-widths at most $0.27$. Bold marks the best deployable arm; \PG
is the own-reserve upper bound; a larger $\varphi$ helps \NR under drift
and on 2026
(\S\ref{sec:eval-q4}).}
\label{tab:main}
\begin{tabular}{@{}llcccccc@{}}
\toprule
snapshot & work & LN & Shaduf & Horcrux & $\Pi_{\mathrm{part}}$
  & \NR $\alpha{=}0.5$ & $\Pi_{\mathrm{glob}}$ \\
\midrule
2021 & skew  & $45.4$ & $60.2$ & $\mathbf{63.8}$ & $57.5$ & $63.2$ & $67.6$ \\
2021 & drift & $35.0$ & $54.5$ & $\mathbf{68.1}$ & $57.0$ & $67.4$ & $76.5$ \\
2023 & skew  & $64.8$ & $79.7$ & $79.5$ & $78.9$ & $\mathbf{80.5}$ & $81.5$ \\
2023 & drift & $56.2$ & $76.5$ & $81.2$ & $79.0$ & $\mathbf{87.0}$ & $91.3$ \\
2026 & skew  & $75.6$ & $82.3$ & $\mathbf{88.3}$ & $82.2$ & $84.8$ & $95.5$ \\
2026 & drift & $66.0$ & $76.8$ & $\mathbf{83.6}$ & $81.6$ & $82.8$ & $84.8$ \\
\bottomrule
\end{tabular}
\end{table}

\begin{table}[t]
\centering\footnotesize
\setlength{\tabcolsep}{4pt}
\caption{What the split forfeits and what \NR recovers, $L{=}20$.
$L_{\mathrm{res}}$ is the share of the pooling gain $G_{\mathrm{glob}} -
G_{\mathrm{LN}}$ that the partition forfeits; $R$ is the share of that loss,
the gap, that \NR recovers, and is fragile where the gap is small. Both are
shares, not success rates: on 2021 under skew pooling adds $22.3$ points,
the partition forfeits $10.1$ of them and \NR at $\alpha{=}0.5$ recovers
$5.7$.}
\label{tab:lres}
\begin{tabular}{@{}llccccc@{}}
\toprule
snapshot & work & pooling gain (pt) & gap (pt) & $L_{\mathrm{res}}$ & $R$, $\alpha{=}0.9$ & $R$, $\alpha{=}0.5$ \\
\midrule
2021 & skew  & $22.3$ & $10.1$ & $45\%$ & $31\%$ & $56\%$ \\
2021 & drift & $41.5$ & $19.5$ & $47\%$ & $25\%$ & $53\%$ \\
2023 & skew  & $16.8$ & $2.7$ & $16\%$ & $43\%$ & $60\%$ \\
2023 & drift & $35.0$ & $12.3$ & $35\%$ & $39\%$ & $65\%$ \\
2026 & skew  & $19.9$ & $13.3$ & $67\%$ & $3\%$  & $19\%$ \\
2026 & drift & $18.8$ & $3.2$ & $17\%$ & $42\%$ & $36\%$ \\
\bottomrule
\end{tabular}
\end{table}

\findingbox{\textbf{Finding.} The partition forfeits $16$ to $67\%$ of the
pooling gain; \NR recovers $19$ to $65\%$ of that loss and stacks positively
on a coin mover in eleven of twelve cells.}
Pooling is worth $16.8$ to $41.5$ points over unpooled channels;
Table~\ref{tab:lres} reads Table~\ref{tab:main} as shares of that gain. \NR recovers least on 2026 under skew, where the loss sits
on nodes with hundreds of channels and half the free reserve cannot cover a
gap that scales with $n$. \NR is ahead of Shaduf in every cell and of
Horcrux in two of six, but the two families meet coordination cost
differently by construction, and they stack.

\begin{figure}[t]
\centering
\includegraphics[width=0.78\textwidth]{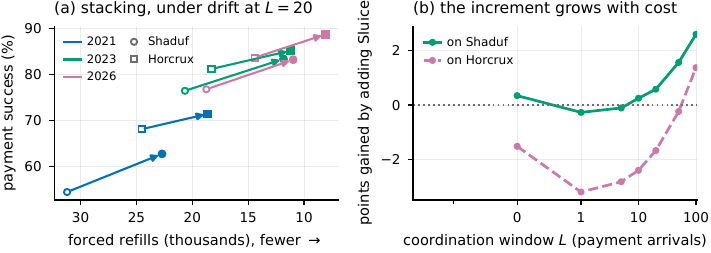}
\caption{\textbf{(a)} Stacking on the success--refill plane under drift,
$L{=}20$; each arrow runs from a coin mover to that mover plus \NR, and
up-and-right is better on both axes. \textbf{(b)} The increment against the
coordination window, 2021 skew.}
\label{fig:stack}
\end{figure}

\paragraph{Composing with the coin movers.}
A coin mover redistributes the balance already inside the channels; \NR
re-deploys a reserve skimmed out of them. Fig.~\ref{fig:stack} shows the
two compose: stacking \NR at $\alpha=0.9$ on Shaduf or Horcrux is positive
in eleven of
twelve cells at $L=20$, by up to $8.3$ points under drift, where one-way flow
drains a whole neighbourhood so a path-local shift finds no source, and
negative only for Horcrux on 2021 under skew, where the mover already
reaches what is needed (all twelve cells in Appendix~\ref{app:sens}).

\subsection{Q2: what pooling costs}
\label{sec:eval-q2}

\begin{figure}[t]
\centering
\includegraphics[width=0.78\textwidth]{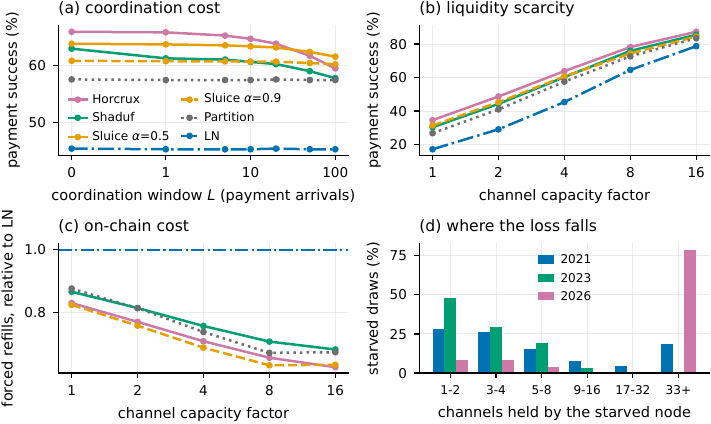}
\caption{Four sweeps on 2021 under skew unless noted. \textbf{(a)} the
coordination window, \textbf{(b)} the capacity factor, \textbf{(c)} forced
refills relative to LN, \textbf{(d)} starved draws by the channel count of
the starved node, all three snapshots.}
\label{fig:cost}
\end{figure}

\findingbox{\textbf{Finding.} \NR loses at most $1.3$ points to coordination
where the coin movers lose up to $10.3$. Its price is its anchors, $2.8$ to
$5.1$ times LN's on-chain bytes or $0.5$ to $1.2$ times incrementally, and a
fee for signers.}

\paragraph{Coordination.}
From $L{=}0$ to $L{=}20$, Shaduf loses $1.6$ to $5.5$ points and Horcrux
$1.9$ to $10.3$, while \NR moves by at most $0.2$ at $\alpha=0.9$ and $1.3$
at $\alpha=0.5$ (Fig.~\ref{fig:cost}a), because most draws never leave the
base. From $p=0.8$ to $0.95$ \NR moves by at most $0.7$ points and Shaduf,
which needs one named party, by up to $2.7$ (Appendix~\ref{app:sens}).

\paragraph{Signers who will not sign.}
A refusing signer is, to the protocol, an absent one. In our runs $47$ to
$90\%$ of pooled nodes never draw and so gain nothing from access. Stressing
availability to $p=0.6$ and $0.4$, \NR falls below the partition in five of
six cells at $p=0.4$. The fee of
\S\ref{sec:signers} is therefore necessary.

\begin{table}[t]
\centering\footnotesize
\setlength{\tabcolsep}{5pt}
\caption{\NR's on-chain cost, $\alpha=0.5$, $L=20$, five seeds: total vbytes
per payment relative to LN, anchors counted with their inputs, and \NR's
gain in payment success over the partition at the same epoch, as the range
over the six cells. Per-cell bytes are in Appendix~\ref{app:sens}.}
\label{tab:onchain}
\begin{tabular}{@{}llcc@{}}
\toprule
anchoring & epoch $\tau_e$ & bytes vs LN & gain over \PP (pt) \\
\midrule
full        & $3{,}500$  & $7.4$--$12.7\times$ & $1.4$ to $12.6$ \\
full        & $14{,}000$ & $2.8$--$5.1\times$  & $0.9$ to $9.1$ \\
full        & $56{,}000$ & $1.1$--$2.1\times$  & $-0.3$ to $2.8$ \\
incremental & $3{,}500$  & $0.7$--$1.4\times$  & $-0.1$ to $4.4$ \\
incremental & $14{,}000$ & $0.5$--$1.2\times$  & $0.2$ to $5.4$ \\
\bottomrule
\end{tabular}
\end{table}

\paragraph{On chain.}
A forced refill is a splice of fresh funds into a starved channel, the
on-chain alternative to every mechanism here, and \NR at $\alpha=0.5$ needs
fewer of them than either coin mover in every cell (Fig.~\ref{fig:cost}c).
Anchors are the rest of the bill (Table~\ref{tab:onchain}). A full anchor
spends and re-creates $n+2$ outputs, about $100(n+2)$ vbytes. Only $0.5$ to
$3.3\%$ of an anchoring node's channels overflow in an epoch, so an
incremental anchor costs a twenty-fifth to a sixth of a full one, but
standing bases hold their headroom and it keeps only part of the gain.
If one counterparty in ten, or four in ten, does not co-sign its re-created
base, success moves by at most $0.14$ or $0.62$ points
(Appendix~\ref{app:sens}).
Nodes with
more than $32$ channels carry $37$ to $46\%$ of the anchor bytes.

\paragraph{Where the loss falls.}
We count a draw as \emph{starved} when its channel's share is spent while the
node's reserve is not, and bin starved draws by the starved node's channel
count (Fig.~\ref{fig:cost}d). On 2026, $78.5\%$ sit on nodes with more than
$32$ channels, against $0.4\%$ on 2023, and under skew the snapshot with the
loss on large nodes has the largest $L_{\mathrm{res}}$. This is
Proposition~\ref{prop:frag} measured: the partition's loss falls on the
routing nodes.

\subsection{Q3: does the bound hold}
\label{sec:eval-q3}

\begin{figure}[t]
\centering
\includegraphics[width=0.66\textwidth]{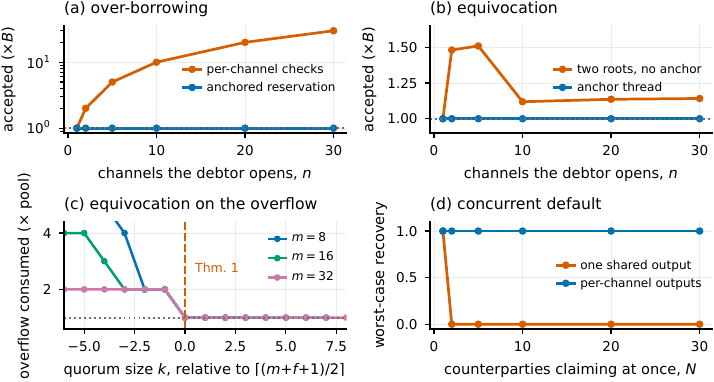}
\caption{The attack matrix. \textbf{(a)} over-borrowing:
per-channel checking alone concedes $n\times$ the reserve, the anchored
reservation does not; \textbf{(b)} equivocation on the
root; \textbf{(c)} equivocation on the overflow against quorum size:
consumption stays at the overflow at the threshold Theorem~\ref{thm:overflow}
requires and reaches twice it one step below, no colluding member shown and
the
same at every $f$ swept; \textbf{(d)} concurrent default.}
\label{fig:attacks}
\end{figure}

\findingbox{\textbf{Finding.} The bound holds at exactly $1.000$ of the
overflow at the quorum threshold; one step below it, an equivocating node
draws twice the overflow.}
Seven adversaries run against the policy code, each swept;
Fig.~\ref{fig:attacks} shows the four that bear on the aggregate bound.
Per-channel checking alone concedes $30\,B$ at $n=30$ where the anchored root
holds $1.00\,B$ (a), and equivocation on the overflow reaches twice it
one step below the threshold and four times further down (c), which is the
intersection argument measured. Appendix~\ref{app:attacks} adds sybil members, which
break a counted quorum and not a weighted one unless $F$ is under-estimated,
and replay after a crash and reuse on two channels, which fail.

\subsection{Q4: the knob}
\label{sec:eval-q4}
\begin{figure}[t]
\centering
\includegraphics[width=0.52\textwidth]{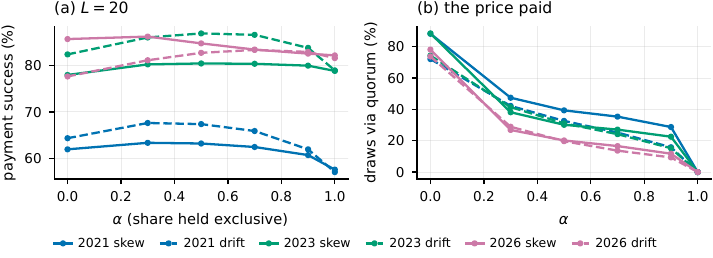}
\caption{Sweeping $\alpha$ at capacity factor $4$. \textbf{(a)} success with
coordination charged, \textbf{(b)} the share of draws needing a quorum.}
\label{fig:alpha}
\end{figure}

\findingbox{\textbf{Finding.} One of $\alpha=0.3$ or $0.5$ is within
$0.7$ points of the per-cell optimum in every cell.}
With coordination free the best setting is
$\alpha = 0$ (Appendix~\ref{app:sens}). Once coordination is charged the optimum moves inside, to $0.3$
or $0.5$ in five cells and $0.7$ in one, because at $\alpha = 0$ up to $88\%$
of draws need a quorum and by $\alpha = 0.3$ under half do
(Fig.~\ref{fig:alpha}). Lowering $\alpha$ per node
with its channel count trails the best fixed $\alpha$ in five of six cells:
the extra quorum traffic on hubs costs more than it returns. A larger
reserve, $\varphi=0.6$, adds $1.3$ to $3.5$ points to \NR under drift and on
2026, so Table~\ref{tab:main} does not show \NR at its best $\varphi$ (nor
the partition, whose optimum under drift is $0.45$ to $0.60$;
Appendix~\ref{app:sens}, \ref{app:space}).

\section{Discussion and Conclusion}
\label{sec:discussion}

\paragraph{Limitations and future work.}
\emph{Assumptions.} \NR does not make the signing set honest: sybil members
cost locked capital, and $F$, the design's main weakness, is each
counterparty's choice. Capacity is verifiable on chain but ownership is not;
membership rules that weigh a channel's age or history are the next step.
\emph{Deployment.} No Lightning implementation carries \NR yet; building one
and testing it on a test network come next. \emph{Incentives.} Safety needs none; access rests on
reciprocity and a fee, neither formalised. \emph{Privacy.} A signer learns
that the node drew on the overflow, when and how much, though not on which
channel, and the base outputs reveal each channel's debt on chain
(Table~\ref{tab:leak}); padding and range proofs~\cite{avarikioti2022hide}
would blunt this. \emph{Evidence.}
Results are simulated on real topology without retries or slow
signers; replaying real payment traces is the natural check.

\paragraph{Conclusion.}
Pooling a node's channels means enforcing one global invariant from many
local views. \NR keeps advance reservation for the common case and adds a
quorum-certified overflow for the rest. Code and data will be made public upon acceptance.

\FloatBarrier
\clearpage
\bibliographystyle{splncs04}
\bibliography{sluice_arxiv}

\clearpage
\appendix

\noindent\textbf{Appendices.} Appendix~\ref{app:ds} specifies \NR's data
structures, algorithms, failure cases and the signing-set rules in full;
Appendix~\ref{app:proofs} proves the lemmas and the theorem of
\S\ref{sec:security}; Appendix~\ref{app:proto} reports the regtest
prototype; Appendix~\ref{app:attacks} gives the quorum availability model
and the attacks not drawn in \S\ref{sec:eval-q3}; Appendix~\ref{app:sens}
collects the sensitivity sweeps; Appendix~\ref{app:ideal} gives the
coordination-free comparison and baseline fidelity; and
Appendix~\ref{app:space} the repairs of \S\ref{sec:space}.

\section{Data Structures and Algorithms}
\label{app:ds}
This appendix states the mechanism of \S\ref{sec:nest} at the level of detail
an implementation needs. Table~\ref{tab:sym} collects the notation and
Table~\ref{tab:leak} what each party learns.

\begin{table}[htbp]
\centering\footnotesize
\setlength{\tabcolsep}{4pt}
\caption{Notation.}
\label{tab:sym}
\begin{tabular}{@{}ll@{}}
\toprule
$n$, $B$, $\varphi$ & channels of the node, reserve, skim fraction \\
$D_i$, $\mathrm{base}_i$, $\mathrm{overflow}$ & drawn on channel $i$, its exclusive share, the shared overflow \\
$\alpha$, $g$, $\tau_e$ & exclusive share of free capacity, leaf group size, epoch length \\
$\delta$, $\delta'$, $C'$ & a draw, its part beyond the base, overflow consumed after it \\
$R_t$, $h_{\mathrm{prev}}$, $b = h(i\|r)$ & epoch root, preceding certificate's hash, channel binding \\
$S_t$, $m$, $k$, $f$ & signing set, size, on-chain count threshold, colluding members \\
$\mathrm{cap}_j$, $W$, $F$ & member capacity, total, adversarial bound (off-chain quorum) \\
$c_j$, $h_j$, $p_j$ & signer's certified total, its hash, pending request \\
$L$, $p$, $L_{\mathrm{res}}$, $R$ & coordination window, availability, loss forfeited, share recovered \\
\bottomrule
\end{tabular}
\end{table}

\begin{table}[htbp]
\centering\footnotesize
\setlength{\tabcolsep}{5pt}
\caption{What a counterparty $w_j$ learns beyond its own channel. The base
path matches Lightning today; the overflow path and the anchor do not.}
\label{tab:leak}
\begin{tabular}{@{}lccc@{}}
\toprule
 & base path & overflow path & anchor \\
\midrule
a single other channel's base & yes$^a$ & no & yes$^b$ \\
a single other channel's debt & no & no & yes$^b$ \\
aggregate base of other subtrees & yes & yes & no \\
that some channel drew on the overflow, and $\delta'$ & no & yes & no \\
cumulative overflow consumption $C'$ & no & yes & no \\
which channel drew & no & no & no \\
the node's channel count $n$, already in gossip & no & no & yes \\
the membership of $S_t$ & no & no & yes \\
\bottomrule
\end{tabular}
\par\smallskip\raggedright\scriptsize $^a$ the sibling leaf at the lowest level of its path. $^b$ from the base outputs' values on chain, since $\mathrm{base}_i - D_i$ is the same for every re-created channel.
\end{table}

\paragraph{State.}
The node $N$ keeps, for the current epoch $t$: the root $R_t$ and the tree that
produced it, the per-channel debts $\{D_i\}$, the bases $\{\mathrm{base}_i\}$,
the overflow's size, the cumulative consumption $C$ and the hash
$h$ of the certificate that set it, and the signing set with its members'
capacities $\mathrm{cap}_j$, their total $W$, and the on-chain count
threshold $k$ used only for settlement.
A counterparty $w_j$ keeps only: $R_t$, its own leaf
$\langle j, \mathrm{base}_j \rangle$ and Merkle path, its own debt $D_j$, the
overflow's size, the signing set $S_t$ with its capacities read from the
anchor, its own bound $F_j$, its salts, and the three durable values
$(c_j, h_j, p_j)$. It learns that some channel drew on the overflow only when
it is asked to endorse the draw; what it learns about other channels is in
Table~\ref{tab:leak}.

\paragraph{The commitment.}
$R_t$ is the root of a Merkle sum tree over the leaves
$\mathrm{leaf}_i = h(\,i \,\|\, \mathrm{base}_i\,)$ paired with the value
$\mathrm{base}_i$; an internal node is
$h(\,v_L \| \mathrm{left} \| v_R \| \mathrm{right}\,)$ carrying
$v = v_L + v_R$. A path of length $\lceil \log_2 n\rceil$ therefore proves both
that $\mathrm{base}_i$ is the committed share of channel $i$ and that the
declared total at the root is the sum of all shares, which is the property
Lemma~\ref{thm:base} uses. The path reveals the value of every sibling
subtree: aggregates of other channels' bases at every level, and one other
channel's base at the lowest. We do not group leaves to hide that one base,
since a member that receives only its group's total cannot check how the
total is split, and the node could then promise the group more than the
total; the bases are on chain in any case (Table~\ref{tab:leak}). The root is committed on chain in an
\texttt{OP\_RETURN} output of the anchor transaction, and the anchor spends the
previous epoch's thread output, so at most one root exists per epoch.

\paragraph{The certificate.}
A certificate is the tuple
$(R_t,\, \delta',\, C',\, h_{\mathrm{prev}},\, b,\, \sigma)$ where $\sigma$
holds signatures over the first five fields, each tagged with the index of
the member that produced it so a verifier can look up its capacity, from
members whose capacity exceeds $(W+F)/2$: the epoch root, the amount
drawn from the overflow, the cumulative overflow consumption after this draw, the hash
of the preceding certificate, and the channel binding $b = h(i \,\|\, r)$ for a
salt $r$ the node draws fresh per certificate and gives to $w_i$ alone. Its
own hash $h(\cdot)$ over those five fields becomes the next certificate's
$h_{\mathrm{prev}}$, so the certificates of one epoch form a hash chain rooted
at $R_t$. $w_i$ accepts the draw only if $\sigma$ verifies against $S_t$ and
$b = h(i \,\|\, r)$, so a certificate serves exactly one channel; the signers
see $b$ and cannot invert it, and $w_i$ keeps $(q, \sigma, r)$ as its claim on
default. At $32$-byte hashes, $8$-byte amounts and $64$-byte Schnorr
signatures, a $k=3$ certificate is $304$\,B and grows by $64$\,B per
additional signer.

\paragraph{Procedures.}
Algorithm~\ref{alg:draw} gives the draw, the endorsement and the
receiver's acceptance, Algorithm~\ref{alg:epoch} the epoch transition, and
Algorithm~\ref{alg:recover} what a signer does after losing its durable
state.

\begin{algorithm}[!t]
\footnotesize
\DontPrintSemicolon
\SetKwInOut{State}{signer state}
\Fn{\textnormal{$\mathrm{Draw}(i,\delta)$} \normalfont{run by the node $N$}}{
  \lIf{$D_i + \delta \le \mathrm{base}_i$}{\Return \textnormal{$w_i$ settles
    locally against $R_t$}}
  $\delta' \gets D_i + \delta - \max(D_i, \mathrm{base}_i)$
    \tcp*{only the new part}
  $C' \gets C + \delta'$;\quad $r \gets$ fresh salt;\quad
  $q \gets (R_t, \delta', C', h_{\mathrm{prev}}, h(i\|r))$;\quad
  $\sigma \gets \langle\,\rangle$\;
  \ForEach{$w_j \in S_t$}{
    send $w_j$ the certificates after its head $h_j$
      \tcp*{catch-up}
    $s \gets \mathrm{Endorse}_j(q)$;\quad
    \lIf{$s \ne \bot$}{append $s$ to $\sigma$}
    \lIf{$\mathrm{cap}(\sigma) > (W + F_i)/2$}{$C \gets C'$; $h_{\mathrm{prev}} \gets h(q)$; broadcast
      $(q,\sigma)$ to $S_t$; \Return $(q,\sigma,r)$ \textnormal{to $w_i$}}
  }
  \Return $\bot$ \tcp*{not enough capacity signed: retry $q$ later, or fall
    back to $\mathrm{base}_i$}
}
\vspace{3pt}
\State{$c_j, h_j$: the last certified total $w_j$ knows and its hash;
$p_j$: the request signed on top of $h_j$ and not yet certified, or $\bot$.
All three durable.}
\Fn{\textnormal{$\mathrm{CatchUp}_j(q', \sigma')$} \normalfont{run by $w_j$ on
  any certificate it is shown}}{
  \lIf{$\mathrm{cap}(\sigma') \le (W+F_j)/2$ \textnormal{over valid
    signatures from $S_t$, or} $q'.h_{\mathrm{prev}} \ne h_j$}{\Return}
  $c_j \gets q'.C'$;\quad $h_j \gets h(q')$;\quad $p_j \gets \bot$\;
}
\Fn{\textnormal{$\mathrm{Endorse}_j(q = (R_t, \delta', C', h_{\mathrm{prev}}, b))$}
   \normalfont{run by $w_j$}}{
  \lIf{$R_t$ \textnormal{is not this epoch's anchored root}}{\Return $\bot$}
  \lIf{$C' > \mathrm{overflow}$ \textnormal{\textbf{or}} $C' \ne c_j + \delta'$}
    {\Return $\bot$}
  \lIf{$h_{\mathrm{prev}} \ne h_j$ \textnormal{\textbf{or}}
    $(p_j \ne \bot \textnormal{ \textbf{and} } p_j \ne q)$}{\Return $\bot$}
  $p_j \gets q$; persist $(c_j,h_j,p_j)$ \textnormal{atomically with the
    check}; \Return $\mathrm{sign}_j(q)$\;
}
\vspace{3pt}
\Fn{\textnormal{$\mathrm{Accept}_i(q, \sigma, r)$} \normalfont{run by the
  served counterparty $w_i$}}{
  \lIf{$\mathrm{cap}(\sigma) \le (W+F_i)/2$ \textnormal{over valid
    signatures from $S_t$}}{\Return $\bot$}
  \lIf{$q.b \ne h(i\|r)$ \textnormal{\textbf{or}} $q.R_t \ne R_t$}{\Return $\bot$}
  $D_i \gets D_i + \delta$;\quad keep $(q,\sigma,r)$
    \tcp*{claim on default}
}
\caption{A draw on channel $i$. The base path returns on the first line and
involves nobody else. $\mathrm{cap}(\sigma)$ is the capacity of the members
whose signatures $\sigma$ carries, and $F_i$ is the served counterparty's own
bound; the node collects against the $F$ it expects $w_i$ to hold, and a
signer catching up applies its own $F_j$, so a node that wants every signer
to follow collects above the largest $F$ in use. A signer
commits to at most one successor of its head before signing and advances its
head only on a certified successor; the served counterparty accepts only a
certificate bound to its own channel.}
\label{alg:draw}
\end{algorithm}

\begin{algorithm}[!t]
\footnotesize
\DontPrintSemicolon
\Fn{\textnormal{$\mathrm{Roll}(t \to t{+}1)$} \normalfont{run by the node $N$}}{
  $\mathrm{re} \gets$ channels whose counterparty co-signs (all, under full
    re-anchoring; those that overflowed, under incremental)\;
  $\mathrm{free}' \gets \max(0,\, B - \sum_{i \notin \mathrm{re}} \max(\mathrm{base}_i, D_i) - \sum_{i \in \mathrm{re}} D_i)$\;
  \lForEach{$i \in \mathrm{re}$}{$\mathrm{base}_i \gets D_i + \alpha\,\mathrm{free}'/|\mathrm{re}|$}
  $\mathrm{overflow} \gets (1-\alpha)\,\mathrm{free}'$\;
  build the Merkle sum tree over $\{\langle i,\mathrm{base}_i\rangle\}$, root
    $R_{t+1}$;\quad $C \gets 0$;\quad $h_{\mathrm{prev}} \gets R_{t+1}$\;
  publish the anchor: it \emph{spends the epoch-$t$ thread output}, the
    epoch-$t$ overflow output with the quorum's signatures and the base
    outputs of $\mathrm{re}$ with their counterparties' signatures, and
    creates the new base outputs, one overflow output, a thread output and
    the \texttt{OP\_RETURN} carrying $R_{t+1}$\;
  \lForEach{channel $i$}{send $w_i$ its leaf, its path, and
    $\mathrm{overflow}$}
}
\vspace{3pt}
\Fn{\textnormal{$\mathrm{Accept}_j(R_{t+1}, \mathrm{leaf}, \pi,
   \mathrm{overflow})$} \normalfont{run by $w_j$}}{
  \lIf{$R_{t+1}$ \textnormal{is not confirmed by an anchor spending epoch $t$'s
    thread output}}{\Return $\bot$}
  \lIf{$\pi$ \textnormal{does not verify, or} $\mathrm{base}_j < D_j$}
    {\Return $\bot$ \tcp*{a base may never fall below outstanding debt}}
  adopt $(R_{t+1}, \mathrm{base}_j, \mathrm{overflow}, S_{t+1},
    \{\mathrm{cap}\}, W)$;\quad
  $c_j \gets 0$;\quad $h_j \gets R_{t+1}$;\quad $p_j \gets \bot$\;
}
\caption{The epoch transition. Debt is grandfathered into the new base, so no
accepted draw is ever invalidated, and the overflow counter resets only against
a root the chain has confirmed.}
\label{alg:epoch}
\end{algorithm}

\begin{algorithm}[!t]
\footnotesize
\DontPrintSemicolon
\Fn{\textnormal{$\mathrm{Recover}_j()$} \normalfont{run by $w_j$ after losing
  $(c_j,h_j,p_j)$}}{
  refuse every endorsement request for the rest of epoch $t$\;
  on the anchor of epoch $t{+}1$, run $\mathrm{Accept}_j$ of
    Algorithm~\ref{alg:epoch} \tcp*{fresh $(0, R_{t+1}, \bot)$}
}
\caption{A signer that lost its durable state waits for the next anchor.
Any recovery within the epoch, from zero, from a stale head, or from the
longest chain its peers hold, can be made to sign a second successor of a
state it already extended (Appendix~\ref{app:fail}), which is the one way an
implementation can silently violate Theorem~\ref{thm:overflow}.}
\label{alg:recover}
\end{algorithm}

\subsection{Epoch transitions, absence, recovery and default}
\label{app:fail}
\emph{Rolling an epoch.}
At the boundary the node computes the new allocation from current debts,
publishes the new anchor and distributes fresh leaves and paths. The anchor's
inputs are the previous thread output, the previous overflow output, and the
base outputs it re-creates: all $n$ under full re-anchoring, or only those of
channels that drew beyond their base under incremental anchoring, since a
base that still covers its debt can stand and its leaf keeps the old outpoint.
Fees are paid from the thread output, which carries the node's spare, and the
node's UTXO count stays at $n+2$ in steady state. Each re-created base
needs its counterparty's co-signature; a counterparty that does not answer
leaves its base standing, as under incremental anchoring, and the node
reclaims it only after $T_2$. Appendix~\ref{app:sens} measures this. Under incremental
anchoring the standing bases keep their value, and the re-created channels
$\mathrm{re}$ get $\mathrm{base}_i = D_i + \alpha\,\mathrm{free}'/|\mathrm{re}|$
and the overflow $(1-\alpha)\,\mathrm{free}'$, with $\mathrm{free}' =
\max(0,\, B - \sum_{\mathrm{stand}} \mathrm{base}_i - \sum_{\mathrm{re}}
D_i)$; this keeps $\sum_i \mathrm{base}_i + \mathrm{overflow} = \max(B,
\sum_i D_i)$ and makes the anchor's inputs cover its outputs. Standing bases
hold their headroom, so a node that wants it back re-creates the channels
with the most of it; a node that drew on nothing keeps its previous anchor.

\emph{Channels that open or close mid-epoch.}
A channel that closes cooperatively settles its base output inside the closing
transaction, paying $D_i$ to the counterparty and the rest to the node; its
leaf stays in $R_t$ with $D_i$ frozen, and it leaves $S_t$ at the next anchor.
A channel opened mid-epoch has no leaf, so no base and no overflow draws,
until the next anchor includes it. The overflow counter resets with the new overflow, and a signer accepts a
reset only against a root that the new anchor confirms -- which is why the
counter cannot be laundered by replaying an old epoch's certificate: the root
check will not match.

\emph{An absent signer.}
Nothing breaks. If the signers who answer hold no more than $(W+F)/2$ of
capacity there is no certificate, and the draw falls back to the base
(Prop.~\ref{prop:failsafe}); the node loses access to the shared portion and
no counterparty loses safety within the epoch. This is the sense in which
the design is fail-safe rather than fail-open.

\emph{A signer that lost its state.}
This is the one place where an implementation can silently break the argument,
so we make the rule explicit and say why nothing weaker works. A signer that
has lost $(c_j,h_j,p_j)$ may already have signed a successor of its last head
and cannot know it. Recovering the head from its peers does not help: the node
may hold a complete certificate $q_1$ that $w_j$ signed and simply not have
released it, so every honest peer's chain still ends at the state below
$q_1$; $w_j$ would adopt that state, sign a successor $q_2$ of it, and the
node would then release $q_1$, giving one state two certified successors. With
$f=0$ and $k=(m+1)/2$ the second quorum is $w_j$ plus the $m-k$ members that
never saw $q_1$, exactly $k$. Asking peers for their pending requests fails
the same way, since the node chooses whom to ask and when. The signer
therefore endorses nothing until the next anchor, whose fresh root $R_{t+1}$
no old request can name; the cost is at most one epoch of that signer's
availability, and the quorum tolerates it as it tolerates any absence.

\emph{A node that withholds.}
A node that stops publishing the overflow state, or publishes an inconsistent
one, is refused by every honest signer and drops to \PP. A node that stops
anchoring altogether leaves the previous epoch's outputs standing, and each
$w_i$ claims its base independently.

\paragraph{Default.}
The base output's cooperative path pays $w_i$ its $D_i$ and returns the rest
to the node; if the node does not cooperate, $w_i$ takes the whole base
output after $T_1$.
If the node defaults, each $w_i$ claims its base output alone, which covers
$D_i$ up to $\mathrm{base}_i$. Overflow debt, $D_i - \mathrm{base}_i$ for the
channels that drew beyond their base, is covered by the overflow output, and
the certificate chain is the ledger of who is owed what: the last certificate
fixes the total consumed, and each creditor holds the certificates that served
it. Settlement is one $k$-of-$m$ transaction paying each overflow creditor its
consumed amount and returning the remainder to the node, which any creditor
may assemble; each creditor reveals its salt so the signers can match its
certificates to its channel, and honest members of $S_t$ sign because the
chain then determines the split. The node cannot assemble a settlement in
its own favour because it holds fewer than $k$ members. If $k$ willing signers cannot be reached, the timelock returns
the whole output to the node after $\Delta'$, and overflow creditors are
exposed to the extent of their overflow debt. That is the one place where
\NR's liveness assumption becomes an exposure, and it is bounded by
$\mathrm{overflow}$ in total and by what each creditor chose to accept beyond
its base.

\subsection{The signing set, in full}
\label{app:signers}
\emph{The range of $F$.} Safety needs $F$ at least the capacity $s$ the node
truly controls in $S_t$; availability needs the capacity it does not
control, $W-s$, to exceed the threshold $(W+F)/2$. The workable range is
therefore $s \le F < W - 2s$, and no $F$ works once $s \ge W/3$, where the
overflow is unusable and the node should run $\alpha=1$.
\emph{Different $F$.} Counterparties may hold different $F_i$: two
certificates accepted under $F_i$ and $F_j$ share capacity above
$(F_i+F_j)/2 \ge s$ (Lemma~\ref{lem:wqi}).
\emph{Count and weight.} On chain the tapleaf enforces a count threshold $k$,
which governs who can assemble the default settlement; the weight check
lives in the signers and the receiver, where draws are decided. The two are
tied by the membership floor: a node with $F$ of capacity holds at most
$F/c_{\min}$ members, so $c_{\min} > 2\max_i F_i/m$ keeps it below any count
majority as well. At $F=0.2W$ the floor is $0.01$ to $0.02$\,BTC for the
median hub on our snapshots and excludes $30$ to $49\%$ of its channels from
$S_t$; it halves with $F$.
\emph{Capacities.} The node takes each $\mathrm{cap}_j$ at anchor time and
commits the vector alongside $R_t$; membership is re-evaluated at each
anchor, where a member whose capacity fell below $c_{\min}$ leaves $S_t$. The
median node with more than $32$ channels holds $1.5$ to $4.1$\,BTC across
its channels on the three snapshots.
\emph{Size.} A FROST threshold signature would shrink a $k=3$ certificate
from $304$ to about $176$\,B (the $112$\,B of fields and one signature) at
the cost of one extra signer round.

\subsection{Costs}
Per epoch the node builds one tree ($O(n)$ hashes) and publishes one
transaction with $n+3$ outputs, $n+2$ of them spendable; each counterparty verifies one path
($O(\log n)$ hashes). Per base-path draw the counterparty does no work beyond
the comparison it already does today, and no message is sent. Per overflow
draw the node sends one request and collects signatures until their
capacity clears the threshold, so the message complexity is $O(m)$ and the latency one round trip; a $3$-signature
certificate is produced and verified in $0.5$\,ms
(Appendix~\ref{app:proto}). Storage per counterparty is one leaf, one path and
three words.

\section{Proofs}
\label{app:proofs}
\paragraph{Lemma~\ref{thm:base}.}
$w_i$ checks $D_i + \delta \le \mathrm{base}_i$ against a Merkle sum path to
$R_t$ (Appendix~\ref{app:ds} gives the tree and the leaf encoding), so accepting more needs a second leaf under the same root, which
collision resistance forbids. Two roots in one epoch are ruled out because the
anchor transaction spends the previous epoch's thread output, so a second root
would double-spend it. The sum in \eqref{eq:base} at $\alpha=1$ gives the
bound. \qed

\paragraph{Lemma~\ref{lem:wqi}.}
For sets $Q_1, Q_2 \subseteq S_t$, $\mathrm{cap}(Q_1 \cap Q_2) \ge
\mathrm{cap}(Q_1) + \mathrm{cap}(Q_2) - W > (F_i+F_j)/2$. The node controls
at most $\min(F_i,F_j)$, so some member of the intersection is honest; with
one $F$ for all receivers this is the bound $F$. \qed

\paragraph{Lemma~\ref{lem:unique}.}
Suppose two certificates carry different cumulative totals and both name the
same predecessor hash. By Lemma~\ref{lem:wqi} their signer sets share an
honest signer $w_j$; with a counted quorum the same holds from
$|Q_1 \cap Q_2| \ge 2k-m \ge f+1$. An honest signer signs a
successor of $h_j$ only when $p_j$ is empty or equals the request, and $p_j$
is written before the signature leaves, so $w_j$ signs at most one successor
of any state, and it signed at most one of the two. Hence no two conflicting
certificates both reach a quorum. Catch-up does not add a case: $w_j$ advances
$h_j$ only on a certificate whose signers hold more than $(W+F_j)/2$ of
capacity, so the state it advances to is itself certified. By induction on certificate depth, every
certified state therefore has a unique certified successor, so the certified
totals form a single chain rather than merely pairwise consistent ones. \qed

\paragraph{Theorem~\ref{thm:overflow}.}
The bases are bounded exactly as in Lemma~\ref{thm:base}, since nesting adds
a path and does not change the committed leaves. By Lemma~\ref{lem:unique}
the certified totals of an epoch form one chain, and every signer checks
$C' \le \mathrm{overflow}$, so the last total in the chain respects it. Each certified $\delta'$ is credited to one channel, since $w_i$
accepts only a certificate whose binding opens to $i$ and the salt is known to
$w_i$ alone, and each certificate carries only the part of a draw beyond the
channel's base or earlier overflow, so the chain's total is the overflow
actually drawn. Hence $\sum_i D_i \le \sum_i \mathrm{base}_i + C \le
\sum_i \mathrm{base}_i + \mathrm{overflow} = \max(B, \sum_i D_i^{t})$, where
$D_i^{t}$ is the debt at the epoch's start; since that debt satisfied
\eqref{eq:bound}, the right-hand side is $B$. \qed


\section{Regtest Prototype}
\label{app:proto}
\paragraph{Prototype.}
To check that this is script that exists rather than script we would like to
exist, we build the outputs and a spending transaction on regtest with
\texttt{python-bitcointx}. The quorum leaf is the standard tapscript multisig
idiom (\texttt{OP\_CHECKSIG}, then $m-1$ \texttt{OP\_CHECKSIGADD}, compared
against $k$) and the timeout leaf uses \texttt{OP\_CHECKSEQUENCEVERIFY}. For
$3$-of-$5$, the threshold at $f=0$, the leaf is $172$\,B and a complete spend is a $521$\,B transaction
with a $429$\,B witness; the leaf and witness grow linearly in $m$, to
$546$ and $1194$\,B at $9$-of-$16$, and are paid only in the default path. Signatures are produced and
verified with \texttt{libsecp256k1}; one from a non-member key, or over a
different cumulative total, fails. A $3$-signature certificate costs
$0.48$ to $0.52$\,ms over four runs on one core of a Xeon Platinum 8375C at 2.9\,GHz, so the
coordination latency of
\S\ref{sec:eval-q2} is round-trips, not cryptography. Every anchor output
uses the BIP~341 unspendable point as its internal key, so no party can
spend it alone by key path.

\emph{Re-anchoring.} A second script builds the base outputs of
\S\ref{sec:nest-epoch}, each with a cooperative 2-of-2 leaf, a claim leaf
for $w_i$ after $T_1=144$ blocks and a reclaim leaf for the node after
$T_2=1008$, and a full re-anchor for $n=5$: the thread by key path, the
overflow through its $3$-of-$5$ quorum leaf and the five bases through their
cooperative leaves, into five new bases, an overflow, a thread and the
\texttt{OP\_RETURN}. All fifteen signatures verify against their sighashes
and a signature from an outsider does not. The transaction is $1{,}100$\,vB for $n=5$.
Through the 2-of-2 script leaf a base input costs $107.8$\,vB; as a MuSig2
key path (BIP~327), the form a deployment would use and the one
Appendix~\ref{app:sens} prices, it costs $57.5$\,vB, so without MuSig2 the
full anchor bytes of Table~\ref{tab:onchain} rise by about half. A default
claim by $w_i$ after $T_1$ is a $146$\,vB transaction. All primitives are
deployed consensus rules (BIP\,341, 342, 112).

\section{Coordination Model, Quorum Availability and Further Attacks}
\label{app:attacks}
Table~\ref{tab:coordmodel} gives what each arm must reach under the one
coordination rule of \S\ref{sec:setup}. The window excludes the channel's
own commitment update, which every arm including plain Lightning performs.

\begin{table}[htbp]
\centering\footnotesize
\setlength{\tabcolsep}{3pt}
\caption{One coordination rule, applied to every arm. The differences are
structural, not fitted. ``Engaged'' means the party is already in the
payment's path. At $p=0.9$ and our set sizes a quorum is more available than
one named party, since any quorum of $S_t$ suffices.}
\label{tab:coordmodel}
\begin{tabular}{@{}llccl@{}}
\toprule
arm & must reach & engaged? & window & answers \\
\midrule
LN & nobody & --- & $0$ & $1$ \\
\PG (idealised) & one shared counter & --- & $0$ & $1$ \\
\NR base & $w_i$, the next hop & yes & $0$ & $1$ \\
Horcrux & upstream neighbour & yes & $L$ & $1$ \\
Shaduf & another channel's peer & no & $L$ & $p$ \\
\NR overflow & any $k$ of $m$ in $S_t$ & no & $L$
  & $\Pr[\mathrm{Bin}(m,p) \ge k]$ \\
\bottomrule
\end{tabular}
\end{table}

The simulator forms quorums by count. Under the capacity-weighted rule a
quorum needs the online members' capacity to exceed $(W+F)/2$; computed from
the actual channel capacities of every node with more than $32$ channels on
the three snapshots, at $p=0.9$ and $F=0.2W$, that quorum is available with
probability $0.993$ to $0.998$ against $1.000$ by count, so the count model
overstates the overflow's availability by under a point, below the seed
noise of the main table; on 2026 the capacities are the announced proxy.

\paragraph{Specification-level attacks.}
Two further adversaries run the exact sequences of Appendix~\ref{app:fail}
against signers implementing Algorithm~\ref{alg:draw}, for
$m$ from $8$ to $33$, $f \in \{0, m/8, m/4\}$ and $k$ at the
threshold. \emph{Replay after a crash}: the node withholds a complete
certificate $q_1$ that $w_j$ signed, $w_j$ loses its state, and the node
requests $q_2$ on the same predecessor before releasing $q_1$. If $w_j$
recovers by adopting the longest chain its peers hold, a natural alternative
rule, the state gets two certified successors whenever $k$ sits exactly at
$(m+f+1)/2$, that is whenever $m+f$ is odd, and is one signature short
otherwise; if $w_j$ endorses nothing until the next anchor it never does.
\emph{Reuse on two channels}: one certificate for $\delta'$ is shown to two
counterparties at the same head. Without the binding $b$ both accept and the
overflow is charged $\delta'$ for $2\delta'$ drawn, in every configuration; with
it the second refuses, in every configuration.

\begin{figure}[htbp]
\centering
\includegraphics[width=0.9\textwidth]{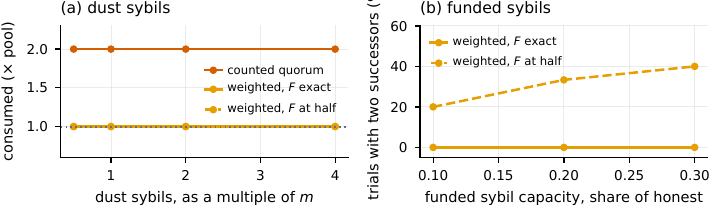}
\caption{Sybil members in the signing set. \textbf{(a)} dust sybils: a
counted quorum is broken from $s = m/2$ on, a weighted one never, whatever
$F$. \textbf{(b)} funded sybils: the weighted quorum holds when $F$ covers
their capacity and breaks in a growing share of trials when $F$ is set at
half of it.}
\label{fig:sybil}
\end{figure}

\paragraph{Sybil members in the signing set.}
The node adds sybil channels to an honest set of $m$ members whose announced
capacities are log-normal, and equivocates as above against two disjoint
honest halves plus all its sybils, for $m \in \{8,16,32\}$. With $s$ dust
sybils, $s$ up to $4m$: a counted quorum, $k = \lceil (m+s+1)/2 \rceil$,
yields two certified successors at every $s \ge m/2$; the capacity-weighted
rule holds at $1.000$ of the overflow at every $s$, and still holds when $F$ is
set at half the sybils' capacity, since dust is dust. With funded sybils
holding $10$, $20$ or $30\%$ of the honest capacity: the weighted rule holds
at $1.000$ when $F$ equals what they hold, and breaks, giving twice the overflow,
in $14$ of $45$ trials when $F$ is set at half of it, whenever the honest
halves split evenly enough. An under-estimated $F$ is therefore the one
setting error that costs safety (Fig.~\ref{fig:sybil}), which is why
\S\ref{sec:signers} tells a cautious counterparty to set it high.

\section{Sensitivity}
\label{app:sens}

\paragraph{Re-anchoring with absent counterparties.}
Re-creating a base output needs its counterparty's co-signature
(\S\ref{sec:nest-epoch}). Table~\ref{tab:cosign} re-runs \NR with each
counterparty answering at re-anchor with probability $0.9$ or $0.6$; a base
whose counterparty does not answer stands at its old value, and the free
capacity is shared among the re-created channels. Success moves by at most
$0.14$ points at $0.9$ and $0.62$ at $0.6$, because a standing base still
covers its channel's debt and only its headroom is misplaced.

\begin{table}[htbp]
\centering\footnotesize
\setlength{\tabcolsep}{5pt}
\caption{\NR $\alpha{=}0.5$ when each counterparty co-signs its re-created
base with probability $q$, $L=20$, $p=0.9$, epoch $10{,}000$, five seeds.}
\label{tab:cosign}
\begin{tabular}{@{}llcccc@{}}
\toprule
snapshot & work & \PP & $q=1$ & $q=0.9$ & $q=0.6$ \\
\midrule
2021 & skew  & $57.5$ & $63.1$ & $63.0$ & $62.6$ \\
2021 & drift & $57.0$ & $67.5$ & $67.4$ & $66.8$ \\
2023 & skew  & $78.9$ & $80.5$ & $80.5$ & $80.4$ \\
2023 & drift & $79.0$ & $87.0$ & $87.0$ & $87.0$ \\
2026 & skew  & $82.2$ & $84.7$ & $84.6$ & $84.1$ \\
2026 & drift & $81.6$ & $82.9$ & $83.0$ & $83.3$ \\
\bottomrule
\end{tabular}
\end{table}

The runs behind \S\ref{sec:eval-q2} and \S\ref{sec:eval-q4}, all at capacity
factor $4$ and $L=20$, five seeds unless stated.

\paragraph{Reserve fraction for \NR.}
The main table inherits $\varphi=0.30$ from the partition's optimum under
skew. Table~\ref{tab:phisl} sweeps $\varphi$ for \NR itself. Under skew the
optimum is $0.30$ or below; under drift, and on 2026 under both workloads, a
larger reserve helps, by $1.3$ to $3.5$ points at $\varphi=0.6$, because the
overflow lets a larger skim be used where the partition could not place it.
Table~\ref{tab:main} is therefore conservative for \NR in four of six cells.

\begin{table}[htbp]
\centering\footnotesize
\setlength{\tabcolsep}{5pt}
\caption{\NR $\alpha{=}0.5$ against the reserve fraction $\varphi$, $L=20$,
five seeds. Bold is the best $\varphi$ in the row; $0.30$ is the main
table's setting.}
\label{tab:phisl}
\begin{tabular}{@{}llcccc@{}}
\toprule
snapshot & work & $\varphi{=}0.15$ & $0.30$ & $0.45$ & $0.60$ \\
\midrule
2021 & skew  & $62.5$ & $\mathbf{63.1}$ & $62.3$ & $60.5$ \\
2021 & drift & $63.1$ & $67.5$ & $69.3$ & $\mathbf{69.9}$ \\
2023 & skew  & $\mathbf{81.2}$ & $80.5$ & $78.8$ & $76.5$ \\
2023 & drift & $84.2$ & $87.0$ & $88.0$ & $\mathbf{88.3}$ \\
2026 & skew  & $83.4$ & $84.7$ & $87.3$ & $\mathbf{88.2}$ \\
2026 & drift & $80.2$ & $82.8$ & $\mathbf{84.2}$ & $\mathbf{84.2}$ \\
\bottomrule
\end{tabular}
\end{table}

\paragraph{The 2026 capacity proxy.}
The 2026 snapshot's capacities are announced HTLC maxima, not funding
amounts. To see how much rests on them, each channel's
capacity is multiplied by log-normal noise of median one and $\sigma = 0.5$
before balances are split, and the core arms re-run. The partition's loss
moves from $67$ to $62\%$ under skew and from $17$ to $14\%$ under drift,
and \NR's recovery from $19$ to $28\%$ and from $36$ to $45\%$: the proxy
shifts the 2026 numbers by up to nine points and none of the orderings.

\paragraph{Plain replay.}
Table~\ref{tab:replay} repeats the main comparison with senders and
receivers drawn uniformly from the trace, without the skew rule. The
partition forfeits $14$ to $71\%$ of the pooling gain and \NR recovers $27$
to $67\%$ of it, ranges similar to the main table's, so the sender-selection
rule is not where the result comes from. Two orderings differ: on 2021 \NR
edges past Horcrux, and on 2023 the partition edges past both coin movers, since uniform traffic produces fewer
large one-way transfers for a mover to fix and less misplacement for a
partition to suffer.

\begin{table}[htbp]
\centering\footnotesize
\setlength{\tabcolsep}{4pt}
\caption{Plain trace replay, uniform sender selection, capacity factor $4$,
$L=20$, five seeds.}
\label{tab:replay}
\begin{tabular}{@{}lcccccccc@{}}
\toprule
snapshot & LN & Shaduf & Horcrux & $\Pi_{\mathrm{part}}$ & \NR $\alpha{=}0.5$ & $\Pi_{\mathrm{glob}}$ & $L_{\mathrm{res}}$ & $R$ \\
\midrule
2021 & $45.5$ & $61.8$ & $65.2$ & $59.9$ & $66.3$ & $71.9$ & $46\%$ & $53\%$ \\
2023 & $68.1$ & $84.6$ & $84.9$ & $85.8$ & $87.7$ & $88.6$ & $14\%$ & $67\%$ \\
2026 & $69.5$ & $76.7$ & $83.3$ & $75.8$ & $80.2$ & $91.7$ & $71\%$ & $27\%$ \\
\bottomrule
\end{tabular}
\end{table}

\paragraph{Stacking and refills.}
Table~\ref{tab:stackfull} gives the stacking increments of \S\ref{sec:eval-q1}
for all twelve cells, and the forced refills of every arm.

\begin{table}[htbp]
\centering\scriptsize
\setlength{\tabcolsep}{2.4pt}
\caption{Stacking \NR ($\alpha=0.9$) on a coin mover, and forced refills
(thousands), $L=20$, ten seeds. $\Delta$ is the stacked arm minus the mover
alone, in points.}
\label{tab:stackfull}
\begin{tabular}{@{}llcccccccccc@{}}
\toprule
& & \multicolumn{3}{c}{Shaduf} & \multicolumn{3}{c}{Horcrux} & \multicolumn{4}{c}{refills (k)} \\
\cmidrule(lr){3-5}\cmidrule(lr){6-8}\cmidrule(lr){9-12}
snapshot & work & alone & $+$\NR & $\Delta$ & alone & $+$\NR & $\Delta$ & LN & Shaduf & Horcrux & \NR \\
\midrule
2021 & skew  & $60.2$ & $60.8$ & $+0.6$ & $63.8$ & $62.1$ & $-1.7$ & $35.5$ & $26.9$ & $25.1$ & $23.0$ \\
2021 & drift & $54.5$ & $62.8$ & $+8.3$ & $68.1$ & $71.5$ & $+3.3$ & $41.2$ & $31.2$ & $24.5$ & $19.9$ \\
2023 & skew  & $79.7$ & $80.5$ & $+0.8$ & $79.5$ & $79.8$ & $+0.3$ & $24.9$ & $15.9$ & $16.0$ & $13.4$ \\
2023 & drift & $76.5$ & $83.3$ & $+6.9$ & $81.2$ & $85.1$ & $+3.9$ & $32.1$ & $20.7$ & $18.3$ & $9.8$ \\
2026 & skew  & $82.3$ & $83.6$ & $+1.3$ & $88.3$ & $89.0$ & $+0.6$ & $18.1$ & $12.7$ & $9.7$ & $9.4$ \\
2026 & drift & $76.8$ & $83.2$ & $+6.4$ & $83.6$ & $88.7$ & $+5.0$ & $27.0$ & $18.7$ & $14.4$ & $11.0$ \\
\bottomrule
\end{tabular}
\end{table}

\paragraph{Signers who will not sign.}
Table~\ref{tab:rational} gives, for each cell, the share of pooled nodes that
never draw in the run, which is the share with no reason but the fee to
sign, and \NR's success when the signing set's effective availability falls
to $0.4$ or $0.6$. At $p=0.6$ \NR at $\alpha=0.5$ still beats the partition in
four of six cells; at $p=0.4$ it falls below the partition in five, because
the base alone is $\alpha$ of what the partition would have allocated. The
loss is bounded by the epoch: a node that sees its quorum fail re-anchors
with $\alpha=1$ and is the partition again.

\begin{table}[htbp]
\centering\footnotesize
\setlength{\tabcolsep}{4pt}
\caption{Nodes that never draw, and \NR's success when only $40$ or $60\%$
of the signing set answers, $L=20$, five seeds. The partition is unaffected
by $p$.}
\label{tab:rational}
\begin{tabular}{@{}llcc@{}cccc@{}}
\toprule
& & never draw & \PP & \multicolumn{2}{c}{$\alpha{=}0.3$} & \multicolumn{2}{c}{$\alpha{=}0.5$} \\
\cmidrule(lr){5-6}\cmidrule(lr){7-8}
snapshot & work & & & $p{=}0.4$ & $0.6$ & $p{=}0.4$ & $0.6$ \\
\midrule
2021 & skew  & $47\%$ & $57.5$ & $52.5$ & $58.3$ & $54.9$ & $59.4$ \\
2021 & drift & $87\%$ & $57.0$ & $51.1$ & $60.9$ & $55.3$ & $62.6$ \\
2023 & skew  & $67\%$ & $78.9$ & $74.3$ & $77.5$ & $76.8$ & $78.8$ \\
2023 & drift & $90\%$ & $79.0$ & $70.4$ & $79.4$ & $75.4$ & $82.2$ \\
2026 & skew  & $64\%$ & $82.2$ & $82.3$ & $85.5$ & $83.0$ & $84.4$ \\
2026 & drift & $84\%$ & $81.6$ & $74.0$ & $78.2$ & $77.8$ & $80.7$ \\
\bottomrule
\end{tabular}
\end{table}

\paragraph{Signer availability.}
Table~\ref{tab:psens} sweeps $p$, the probability an off-path party answers,
for the two arms it touches. Shaduf needs one named counterparty and pays
$p$ directly; \NR needs any $k$ of $m$ and pays a binomial tail. Averaged over
the six cells, Shaduf gains $1.9$ points from $p=0.8$ to $0.95$ and \NR at
$\alpha=0.5$ gains $0.4$.

\begin{table}[htbp]
\centering\footnotesize
\setlength{\tabcolsep}{4pt}
\caption{Payment success against signer availability $p$ at $L=20$, five
seeds. The $p=0.9$ column is the main table's setting.}
\label{tab:psens}
\setlength{\tabcolsep}{3pt}
\begin{tabular}{@{}llccccccccc@{}}
\toprule
& & \multicolumn{3}{c}{Shaduf} & \multicolumn{3}{c}{\NR $\alpha{=}0.9$}
  & \multicolumn{3}{c}{\NR $\alpha{=}0.5$} \\
\cmidrule(lr){3-5}\cmidrule(lr){6-8}\cmidrule(lr){9-11}
snapshot & work & $0.8$ & $0.9$ & $0.95$ & $0.8$ & $0.9$ & $0.95$
  & $0.8$ & $0.9$ & $0.95$ \\
\midrule
2021 & skew  & 58.7 & 60.2 & 60.9 & 60.5 & 60.7 & 60.6 & 62.7 & 63.2 & 63.2 \\
2021 & drift & 53.1 & 54.5 & 55.2 & 61.9 & 61.9 & 62.0 & 66.9 & 67.4 & 67.5 \\
2023 & skew  & 78.2 & 79.7 & 80.4 & 79.9 & 80.0 & 80.0 & 80.3 & 80.5 & 80.5 \\
2023 & drift & 74.7 & 76.5 & 77.4 & 83.7 & 83.8 & 83.9 & 86.4 & 87.0 & 87.1 \\
2026 & skew  & 81.7 & 82.3 & 82.5 & 82.5 & 82.6 & 82.5 & 84.7 & 84.8 & 84.7 \\
2026 & drift & 76.1 & 76.8 & 77.3 & 83.0 & 83.0 & 83.0 & 82.6 & 82.8 & 82.9 \\
\bottomrule
\end{tabular}
\end{table}

\paragraph{Adaptive $\alpha$.}
\S\ref{sec:eval-q2} finds the partition's loss on high-degree nodes, which
suggests lowering $\alpha$ there. Table~\ref{tab:adapt} tests two step
policies in which each node picks $\alpha$ from its own channel count against
fixed $\alpha$. Policy A trails the best fixed setting in five of six cells
and policy B in all six, and
the one that lowers $\alpha$ most on hubs trails by most: the share of draws
needing a quorum rises from $20$--$39\%$ to $22$--$48\%$, and at $L=20$ that
costs more than the released capacity returns. The exception is 2026 under
skew, where $78.5\%$ of the loss sits on hubs and the adaptive policy leads
by $0.2$. The knob is a property of the network's coordination cost, not of
the node.

\begin{table}[htbp]
\centering\footnotesize
\setlength{\tabcolsep}{4pt}
\caption{Adaptive $\alpha(n)$ against fixed $\alpha$, $L=20$. A: $\alpha=0.8$
for $n\le 8$, $0.5$ for $9\le n\le 32$, $0.2$ above; B: $0.7/0.5/0.3$.
$\Delta$ is against the best fixed setting in the row.}
\label{tab:adapt}
\begin{tabular}{@{}llccccccc@{}}
\toprule
snapshot & work & $\alpha{=}0.3$ & $0.5$ & $0.7$ & A & $\Delta_A$ & B & $\Delta_B$ \\
\midrule
2021 & skew  & 63.2 & 63.1 & 62.4 & 62.3 & $-0.9$ & 62.7 & $-0.6$ \\
2021 & drift & 67.7 & 67.4 & 66.0 & 66.7 & $-1.0$ & 67.3 & $-0.4$ \\
2023 & skew  & 80.3 & 80.5 & 80.4 & 79.4 & $-1.1$ & 79.9 & $-0.6$ \\
2023 & drift & 86.1 & 87.0 & 86.7 & 85.0 & $-2.0$ & 85.9 & $-1.1$ \\
2026 & skew  & 86.2 & 84.7 & 83.4 & 86.5 & $+0.2$ & 86.1 & $-0.2$ \\
2026 & drift & 81.3 & 82.9 & 83.5 & 79.9 & $-3.5$ & 81.1 & $-2.4$ \\
\bottomrule
\end{tabular}
\end{table}

\begin{figure}[htbp]
\centering
\includegraphics[width=0.9\textwidth]{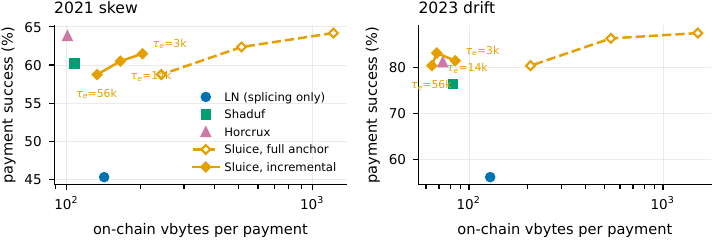}
\caption{Payment success against total on-chain vbytes per payment, $L=20$,
anchors counted with their inputs. \NR is shown with full and incremental
anchoring at epochs of $3{,}500$, $14{,}000$ and $56{,}000$ payments, each
from its own runs; LN and the coin movers anchor nothing.}
\label{fig:onchain}
\end{figure}

\paragraph{On-chain bytes.}
Fig.~\ref{fig:onchain} charges every arm its whole on-chain footprint per
payment: refills at $200$ vbytes each, a splice with one input and two
Taproot outputs, plus for the pooled arms the anchors of every node that drew
during the epoch, since a node with nothing to re-allocate keeps its previous
anchor. An anchor is counted with its inputs: $60 + 100.5\,(c+2)$ vbytes for
$c$ re-created base outputs, each a $57.5$-vbyte key-path input plus a
$43$-vbyte output, which assumes the cooperative path is a MuSig2 key path
(Appendix~\ref{app:proto}). Under full re-anchoring $c=n$; under incremental anchoring
$c$ is the number of channels that drew beyond their base in the epoch, which
we count in the same runs: $0.5$ to $3.3\%$ of $n$, so the incremental anchor
costs $4$ to $16\%$ of the full one. Fee rate multiplies both axes equally, so
the comparison holds at any fee. Table~\ref{tab:onchainfull} gives the
totals per cell relative to LN, each from its own runs. The incremental form
is simulated in its own right, with standing bases kept and $\mathrm{free}'$
shared among the re-created channels: its success is $0.6$ to $3.7$ points
below the full anchor's at $\tau_e = 14{,}000$ and $0.7$ to $9.9$ at $3{,}500$,
because headroom left standing is capacity the overflow cannot reach. Fig.~\ref{fig:onchain} draws the incremental line from these runs.

\begin{table}[htbp]
\centering\footnotesize
\setlength{\tabcolsep}{4pt}
\caption{On-chain vbytes per payment of \NR $\alpha{=}0.5$ relative to LN,
anchors counted with inputs, $L=20$, five seeds. LN's own footprint is
$72$ to $165$ vbytes per payment.}
\label{tab:onchainfull}
\begin{tabular}{@{}llcccccc@{}}
\toprule
& & \multicolumn{3}{c}{full anchor, $\tau_e=$} & \multicolumn{3}{c}{incremental, $\tau_e=$} \\
\cmidrule(lr){3-5}\cmidrule(lr){6-8}
snapshot & work & $3500$ & $14000$ & $56000$ & $3500$ & $14000$ & $56000$ \\
\midrule
2021 & skew  & $8.6$ & $3.6$ & $1.7$ & $1.4$ & $1.2$ & $0.9$ \\
2021 & drift & $7.4$ & $2.8$ & $1.3$ & $0.8$ & $0.7$ & $0.7$ \\
2023 & skew  & $12.7$ & $5.1$ & $2.1$ & $1.4$ & $1.1$ & $0.8$ \\
2023 & drift & $11.8$ & $4.2$ & $1.6$ & $0.7$ & $0.5$ & $0.5$ \\
2026 & skew  & $7.7$ & $3.2$ & $1.5$ & $1.1$ & $0.9$ & $0.8$ \\
2026 & drift & $7.6$ & $2.8$ & $1.1$ & $0.7$ & $0.6$ & $0.5$ \\
\bottomrule
\end{tabular}
\end{table}

Lengthening the epoch from $3{,}500$ to $56{,}000$ costs \NR most of its gain,
$64.2$ to $58.8$ on 2021 under skew and $87.5$ to $80.4$ on 2023 under drift,
because outstanding debt is grandfathered and the overflow is replenished only
at the boundary; the partition, which has no overflow to replenish, moves by
at most $1.9$ points over the same range (Table~\ref{tab:space}). At $56{,}000$
\NR still gains $0.7$ to $2.8$ points over the partition on 2021 and 2023, and
falls just below it on 2026. The coin movers
spend fewer bytes than LN because they prevent refills and anchor nothing.
The anchor bytes follow channel count, so nodes with more than $32$ channels
account for $37$, $41$ and $46\%$ of them on the three snapshots, and a
deployment in which only such nodes pool would pay only that share.

\section{The Idealised Comparison, and Baseline Fidelity}
\label{app:ideal}
Table~\ref{tab:ideal} repeats Table~\ref{tab:main} with coordination free, the
condition the released simulators report; the fidelity check below uses it.
With coordination free, Horcrux leads Shaduf and \NR in four of six cells
(Shaduf leads on 2023 under skew, \NR on 2023 under drift) and on 2021 and
2026 under drift exceeds \PG, $77.3$ against
$76.5$ and $94.0$ against $84.8$, as does Shaduf on 2023 under skew, $82.9$
against $81.5$: \PG bounds what re-deploying the node's own
reserve can achieve, whereas a coin mover also reaches the counterparty's
balance along the path. The ordering reverses once coordination is charged,
because that extra resource has to be reached.

\begin{table}[htbp]
\centering\footnotesize
\setlength{\tabcolsep}{3.5pt}
\caption{Payment success at capacity factor $4$ with coordination free
($L{=}0$), ten seeds, $95\%$ half-widths at most $0.27$.}
\label{tab:ideal}
\begin{tabular}{@{}llcccccc@{}}
\toprule
snapshot & work & LN & Shaduf & Horcrux & $\Pi_{\mathrm{part}}$
  & \NR $\alpha{=}0.5$ & $\Pi_{\mathrm{glob}}$ \\
\midrule
2021 & skew  & $45.4$ & $63.0$ & $\mathbf{65.9}$ & $57.5$ & $63.8$ & $67.6$ \\
2021 & drift & $35.0$ & $59.7$ & $\mathbf{77.3}$ & $57.0$ & $67.6$ & $76.5$ \\
2023 & skew  & $64.8$ & $\mathbf{82.9}$ & $81.5$ & $78.9$ & $81.0$ & $81.5$ \\
2023 & drift & $56.2$ & $81.9$ & $87.4$ & $79.0$ & $\mathbf{88.0}$ & $91.3$ \\
2026 & skew  & $75.6$ & $83.9$ & $\mathbf{95.0}$ & $82.2$ & $84.8$ & $95.5$ \\
2026 & drift & $66.0$ & $81.4$ & $\mathbf{94.0}$ & $81.6$ & $84.1$ & $84.8$ \\
\bottomrule
\end{tabular}
\end{table}

\FloatBarrier
\paragraph{Baseline fidelity.}
\label{app:fidelity}
The coin-moving mechanism code is migrated from the released simulators of
the original work~\cite{ge2022shaduf,tian2024horcrux}. On the 2021 snapshot
they ship, at matched capacity and traffic ($\mathrm{cr}=2$, skew $4$, three
seeds), the released code gives LN $.2718$, Shaduf $.4672$ and Horcrux
$.4591$; ours gives $.2939$, $.4849$ and $.5010$. Normalised by their own LN
baseline the ratios are $1.719$ and $1.689$ against $1.650$ and $1.705$, so
both mechanisms agree within about $4\%$. The level gap has two sources: the payment
trace is a 2021 rather than a 2020 on-chain sample, and our simulator gives
every node an on-chain refill path when a hop cannot be served, where in the
originals the payment simply fails. Refills raise every arm, including LN,
which is why we report refill counts alongside success throughout.

\section{The Repairs of \S\ref{sec:space}, in Full}
\label{app:space}
The sweeps behind \S\ref{sec:space}, in the setup of Table~\ref{tab:main}:
all three snapshots, both workloads, capacity factor $4$, coordination
charged at $L=20$ and $p=0.9$, $\varphi=0.30$ unless swept, $50{,}000$
measured payments. Table~\ref{tab:space} gives one row per cell.

\emph{Re-anchoring more often.} Over a $32$-fold range of epoch length, from
$1{,}750$ to $56{,}000$ payments, the partition's success is nearly flat: the
range is at most $17\%$ of its loss, and under $10\%$ in three of six
cells. The cost is misplacement, not staleness.
\emph{Enlarging the reserve.} The partition has an interior optimum in
every cell, $\varphi=0.30$ under skew and $0.45$ to $0.60$ under drift, and
falls steeply beyond it: on 2021 under skew from $57.4$ at $\varphi=0.30$ to
$48.5$ at $0.90$. In five of six cells the partition at its best $\varphi$
sits below the global check at its worst.
\emph{Allocating by predicted need.} A split proportional to last epoch's
demand loses to the even split in five of six cells, and concentrating the
reserve on last epoch's one or three shortest channels loses $8$ to $28$
points, because last epoch's deficit barely predicts this one's.
\emph{Optimistic accounting.} Each creditor checks the node's limit against
only the debt it can see. An under-reporting node reaches $2.9$ to $111$
times the reserve, and the worst placed creditor recovers $0.9$ to $26\%$ on
default. Shortening the reconciliation window tenfold changes nothing,
because the exposure is set by how many counterparties can be told the same
lie at once. The unsafe variant has the highest success, which is what it
buys by exceeding the bound.

\begin{table}[htbp]
\centering\footnotesize
\setlength{\tabcolsep}{3.5pt}
\caption{The repairs, one row per cell. Gap is $G_{\mathrm{glob}} -
G_{\mathrm{part}}$ in points; epoch range is the partition's success range
over $\tau_e$ from $1{,}750$ to $56{,}000$ as a share of the gap; the
allocation columns are success minus the even split's, in points; the last
two are optimistic accounting's worst excess over the reserve and the worst
placed creditor's recovery. Five to ten seeds.}
\label{tab:space}
\label{tab:epoch}
\label{tab:alloc}
\label{tab:optimistic}
\begin{tabular}{@{}llcccccccc@{}}
\toprule
& & & epoch & \multicolumn{3}{c}{allocation vs even} & \multicolumn{2}{c}{optimistic} \\
\cmidrule(lr){5-7}\cmidrule(lr){8-9}
snapshot & work & gap & range & by demand & top-$1$ & top-$3$ & excess & recovery \\
\midrule
2021 & skew  & $10.1$ & $5.7\%$  & $-2.5$ & $-17.4$ & $-16.3$ & $111.4\times$ & $0.9\%$ \\
2021 & drift & $19.5$ & $9.0\%$  & $-4.4$ & $-26.3$ & $-25.0$ & $97.0\times$  & $1.0\%$ \\
2023 & skew  & $2.7$  & $10.1\%$  & $-1.8$ & $-18.8$ & $-18.0$ & $4.3\times$   & $19.1\%$ \\
2023 & drift & $12.3$ & $15.3\%$ & $-6.4$ & $-28.2$ & $-27.0$ & $8.5\times$   & $10.5\%$ \\
2026 & skew  & $13.3$ & $0.8\%$  & $+0.3$ & $-8.8$  & $-8.0$  & $2.9\times$   & $26.1\%$ \\
2026 & drift & $3.2$  & $17.4\%$ & $-2.0$ & $-19.5$ & $-18.4$ & $26.6\times$  & $3.6\%$ \\
\bottomrule
\end{tabular}
\end{table}


\end{document}